%% file: main.tex
\documentclass[iop,apj,twocolumn,numberedappendix,appendixfloats,twocolappendix]{openjournal}
\usepackage{amsmath}
\usepackage{amsthm}
\usepackage{amssymb}
\usepackage{amsfonts}
\usepackage{physics}
\usepackage{natbib}
\usepackage{xspace}
\usepackage[table]{xcolor} 
\usepackage[normalem]{ulem}
\definecolor{linkcolor}{rgb}{0.16,0.18,0.47}
\usepackage{orcidlink}
\usepackage{hyperref}
\hypersetup{
    unicode, 
    colorlinks=true,
    linkcolor=linkcolor,
    citecolor=linkcolor,
    filecolor=linkcolor,
    urlcolor=linkcolor,
}

\newcommand{\orcidauthor}[3]{\author{\href{http://orcid.org/#1}{#2\hskip2pt\protect\includegraphics[width=9pt]{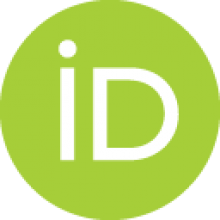}}\,$^{#3}$}}

\newcommand{\RN}[1]{\textup{\uppercase\expandafter{\romannumeral#1}}}
\newcommand{\feh}{\rm{[Fe/H]}}
\newcommand{\ch}{\rm{[C/H]}}
\newcommand{\cfe}{\rm{[C/Fe]}}

\newcommand{\rproc}{\mbox{{\it r}-process}\xspace}

\newcommand{\sproc}{\mbox{{\it s}-process}\xspace}
\newcommand{\iproc}{\mbox{{\it i}-process}\xspace}

\renewcommand{\arraystretch}{1.4}

\begin{document}

%% Title.
\title{Carbon Abundances in Metal-Poor Stars Reveal Distinct Galaxy and\\ Star Formation Pathways in the Early Universe \vspace{-1.5cm}}
\shorttitle{Metal-Poor Star Carbon Abundances Reveal Galaxy and Star Formation Pathways}
\shortauthors{Yelland et al.}

%% Authors.
\orcidauthor{0000-0002-1462-0265}{Alexander Yelland}{1}
\orcidauthor{0000-0002-2139-7145}{Anna Frebel}{1}
\orcidauthor{0000-0002-4669-9967}{Xiaowei Ou}{2,3}
\orcidauthor{0000-0002-7332-2751}{Sarah Hughes}{1}
\orcidauthor{0000-0001-9178-3992}{Mohammad K. Mardini}{4}

%% Affiliations.
\affiliation{$^{1}$ Department of Physics and Kavli Institute for Astrophysics and Space Research, Massachusetts Institute of Technology, 77 Massachusetts Avenue, Cambridge, MA 02139, USA}
\affiliation{$^{2}$ Department of Astronomy, University of Virginia, 530 McCormick Rd, Charlottesville, VA 22904, USA}
\affiliation{$^{3}$ The NSF-Simons AI Institute for Cosmic Origins, USA}
\affiliation{$^{4}$ Physics Department, Faculty of Science, Al Balqa Applied University, 19117 Salt, Jordan}

%% Emails.
% \email{ayelland@mit.edu}
% \email{afrebel@mit.edu}
% \email{xwou@virginia.edu}
% \email{slhughes@mit.edu}
% \email{mmardini@mit.edu}

%%%%%%%%%%%%%%%%%%%%%%%%%%%%%% BODY OF PAPER %%%%%%%%%%%%%%%%%%%%%%%%%%%%%%%%%%%

%%%%%%%%%%%%%%%%%%%%%%%%%%%%%%%%%%%%%%%%%%%%%%%%%%%%
%% Abstract.
\begin{abstract}
Carbon-enhanced metal-poor (CEMP; with $\rm{[Fe/H]} \le -2.0$ and $\rm{[C/Fe]} \ge 0.7$) stars preserve information about early chemical enrichment, low-mass star formation, and the hierarchical assembly of galaxies.
In this study, we have compiled an extensive literature sample of 1045 stellar carbon abundances spanning the metal-poor Milky Way halo (437 stars), 21 ultra-faint dwarf galaxies (UFDs; 109 stars), eight classical dwarf spheroidal galaxies (260 stars), three accreted dwarf galaxies (90 stars), the Small Accreted Stellar Systems (SASS; 77 stars), and eleven stellar streams (72 stars).
We establish the fractions of CEMP stars for each of these systems and categories.
Generally, the low-mass UFDs possess the high fractions at low metallicities, whereas the more massive classical dwarf galaxies have relatively few CEMP stars.
This behavior reveals a new low-metallicity Magnitude ($M_{\rm V}$)--CEMP Fraction relation across the dwarf satellite galaxy population.
The high CEMP fractions in surviving UFDs suggest their enrichment was dominated by faint supernovae, as higher energy input would likely have quenched star production.
The low CEMP fractions in classical dwarfs imply predominantly in situ formation rather than assembly from smaller systems.
Using $\rm{[C/H]}$ abundances, we also probe early low-mass star formation. 
Eight stars lie within or near the theoretical ``forbidden zone'', indicating that dust-induced cooling, alongside fine-structure line cooling, contributed to early star formation.
These rare dust-cooled stars may have formed in UFD-like systems that did not survive.
Overall, the metal-poor Milky Way halo appears to have assembled from many different dwarf galaxies, with CEMP halo stars being contributed primarily by early UFD-like systems and non-CEMP halo stars primarily by intermediate-sized halos that later formed classical dwarfs.
\end{abstract}

%% AAS Keywords (https://astrothesaurus.org)

%%%%%%%%%%%%%%%%%%%%%%%%%%%%%%%%%%%%%%%%%%%%%%%%%%%%
\section{Introduction} \label{sec:intro}

Carbon is one of the most abundant elements in the universe and can be readily measured in stars across a wide range of galactic environments.
In ancient metal-poor stars, surface chemical abundances preserve the nucleosynthetic signatures of their progenitor sources \citep{Frebel2015a}.
As such, very metal-poor stars (\feh\ $< -2$)\footnote{
    The chemical abundance of an element ($\rm{X}$) can be defined by $A(\rm{X}) \equiv \log_{10}\varepsilon (\rm{X}) = \log_{10}\left(N_{\rm{X}}/N_{\rm{H}}\right) + 12$, where $N_{\rm{X}}$ and $N_{\rm{H}}$ are the number densities of the given element and hydrogen. The abundance ratio is defined relative to the absolute solar chemical abundances \citep{Asplund2009a}: $\rm{[X/H]} = A(\rm{X})_{*} - A(\rm{X})_{\odot}$.
} 
serve as ideal probes of carbon production, star formation, and galaxy formation in the early universe, as they retain relatively unaltered records of the enrichment processes that shaped their natal gas clouds where the star formed. 
In this context, carbon abundances in metal-poor halo stars have been extensively investigated for nearly three decades \citep{Rossi1999a, Lucatello2005a, Frebel2006b, Aoki2007a, Carollo2012a, Aoki2013a, HansenT2015a, Placco2014c, Arentsen2022a, Farouqi2025a}.

Large carbon abundances have been found among almost all of the most iron-poor stars with \feh\ $<-4.5$ (e.g., \citealt{Frebel2015b, Aguado2022a, Frebel2008a, Nordlander2019a, Keller2014a}), with many of these stars having values in excess of \cfe\ $>+2$. 
Excluding metal-poor stars with known $s$-process and $i$-process signatures (whose large carbon over-abundances originate from binary mass-transfer from an erstwhile companion and the intermediate neutron-capture process, respectively), there exists a large number of true carbon-enhanced metal-poor (CEMP) stars with \cfe\ $\ge 0.7$ \citep{Aoki2007a} that appear to have formed from very carbon-rich gas in the early universe.
Historically, the fraction of CEMP stars in the halo was found to be 20\% for stars with \feh\ $<-2.0$, 43\% with \feh\ $<-3.0$, and 81\% with \feh\ $<-4.0$ \citep{Placco2014c}. 
This showed the ubiquity of large \cfe\ ratios and a significant carbon production by the first generation(s) of stars.

In the early universe, the metal-free Population\,III (Pop\,III) stars are thought to have been very massive, with $M \sim 10 - 100 \, M_{\odot}$ \citep{Bromm2013a} and varying explosion energies. 
Lower-energy, ``faint'' supernovae (e.g., \citealt{Umeda2005a, Iwamoto2005a, Heger2010a, Nomoto2013a, Ishigaki2014b, Tominaga2014a}) are predicted to produce large amounts of carbon while releasing comparably less iron. 
This occurs via a ``mixing and fallback'' mechanism where the innermost layers, including iron, fall back onto the nascent black hole rather than getting expelled, thus leading to large \cfe\ yields. 
Similarly, aspherical explosions also produce large \cfe\ ratios, in addition to large zinc abundances \citep{Tominaga2009a, Ezzeddine2019a}.
Another proposed mechanisms for early carbon enhancement is massive rotating Pop\,III stars.
Due to rotationally induced mixing, large \cfe\ yields are possible already in the pre-explosion phase as well as during the final explosion (e.g., \citealt{Meynet2006a, Choplin2018a}).
At later times, the increasingly common asymptotic giant branch (AGB) stars typically release significant amounts of CNO elements through stellar winds which adds to the chemical inventory. 
The signatures of these early nucleosynthetic processes should be imprinted in the chemical abundances of the low-mass Population\,II (Pop\,II) stars accessible to observation today.

Within the context of hierarchical galaxy assembly, the Milky Way halo is thought to have formed from a collection of smaller building blocks, including systems similar to the surviving satellite dwarf galaxies \citep{Deason2016a, Deason2023a, Fattahi2020b, Kruijssen2020b, Koutsouridou2026a}.
One might therefore expect metal-poor halo stars and dwarf galaxy stars to generally exhibit similar chemical abundance patterns. 
This appears to be the case in the ultra-faint dwarf galaxies, the ``small accreted stellar systems'', and the stellar streams, where the fraction of CEMP stars at \feh\ $\leq -2.0$ is broadly comparable to that of the halo (e.g., \citealt{Ji2020a}).
In contrast, across essentially all of the more massive classical dwarf spheroidal galaxies and accreted systems (e.g., Atari Disk, Gaia-Sausage/Enceladus, and the Large Magellanic Cloud), the relative fraction of CEMP stars has been known to be lower \citep{Sestito2024d, Chiti2018b, Chiti2024a, Starkenburg2013a, Skuladottir2015a, Lucchesi2024a}, but insufficiently quantified. 
In fact, there is a growing discrepancy in the behavior of stellar carbon abundances among different types of dwarf galaxies.
This has been explored for individual systems theoretically \citep{Salvadori2015a, Koutsouridou2026a} and observationally \citep{Lucchesi2024a} in recent years; however, a systematic observational study across the full range of dwarf galaxies and accreted systems has yet to be done simultaneously within the context of early star formation history and hierarchical galaxy assembly.

Therefore, in this study, we compare stellar carbon enhancement fractions across a range of Milky Way satellites and substructures (including the ultra-faint dwarf galaxies, classical dwarf spheroidal galaxies, accreted dwarf galaxies, and stellar streams) to better understanding the environments and conditions during the onset of galactic chemical enrichment and carbon production by the earliest supernovae.
In doing so, we aim to constrain how early environments and supernovae may have shaped the formation of the first galaxies and subsequent galaxy formation.

%%%%%%%%%%%%%%%%%%%%%%%%%%%%%%%%%%%%%%%%%%%%%%%%%%%%
\section{Literature abundance compilation} \label{sec:lit_prep}

\input{tables/tab1_sysproperties}
\label{tab:sysproperties}

For this study, we have compiled a large number of literature data samples to assess and compare stellar carbon abundances of metal-poor stars in the Milky Way halo to those in ultra-faint dwarf galaxies (UFDs), classical dwarf spheroidal galaxies (CDW, dSphs), accreted dwarf galaxies (ADWs), and stellar streams (SS).
We also consider a sample of ``small accreted stellar systems'' (SASS) stars \citep{Andales2024a, Hughes2026a}, which are included in the category of accreted dwarf galaxy systems.

To construct our sample, we use high-resolution spectroscopic data gathered from JINAbase \citep{Abohalima2018a}, another available compilation of stellar data\footnote{https://github.com/alexji/alexmods.git}, as well as data from many other individual literature sources. 
We generally prioritize measurements based on high-resolution spectroscopy over any available medium-resolution results ($R \gtrsim 2000$).
Systems with available medium-resolution data include e.g., Segue~1, Bootes~I, Sagittarius and Sculptor.
We do not include data with lower resolution.
Additionally, we have chosen to use the abundances determined when assuming local thermodynamic equilibrium (LTE) for consistency across literature data.
The final sample includes stars from twenty-one ultra-faint dwarf galaxies, eight classical dwarf spheroidal galaxies, three accreted dwarf galaxies (in addition to the collection of SASS stars), and eleven stellar streams, all listed in Table~\ref{tab:sysproperties}. 
We also provide basic dwarf galaxy properties such as absolute $V$-band magnitude, stellar mass, sample size, and the CEMP star fractions (see Section~\ref{sec:cempfractions}).

%=================================================
\subsection{Sample preparation}

% metal-poor restriction
To assess early carbon production, our goal is to limit our sample of stars to those formed soon after the first carbon enrichments occurred in the early universe. 
Thus, our sample only includes stars with \feh\ $\le -2.0$.
This metallicity is also below that of the typical ``$\alpha$-knee'' observed in dwarf galaxies \citep{Tolstoy2009a}, which marks the onset of Type\,Ia supernovae and their subsequent dominance of iron production over the Type\,II supernovae, of which occur on shorter-timescales.
Stars below the $\alpha$-knee likely formed during the earliest phases of a galaxy's chemical evolution, making them particularly valuable tracers of the nucleosynthetic yields of the first generations of massive stars.

% carbon band
To ensure consistency when comparing carbon abundances, we only adopt literature carbon abundances derived from the CH $g$-band feature (around $4313\,${\AA}), rather than from the carbon atomic lines or the C$_{2}$ Swan bands, both of which are used less commonly and can yield abundances offset from CH-based measurements.
As such, the sample is naturally limited to the stars across dwarf galaxies and systems with blue optical data available, given that much of the existing spectroscopic literature data only covers the red region (e.g., VLT/FLAMES data for Sculptor, \citealt{Hill2019a}).

\begin{figure*}[p]
    \centering
    \includegraphics[width=6.7in]{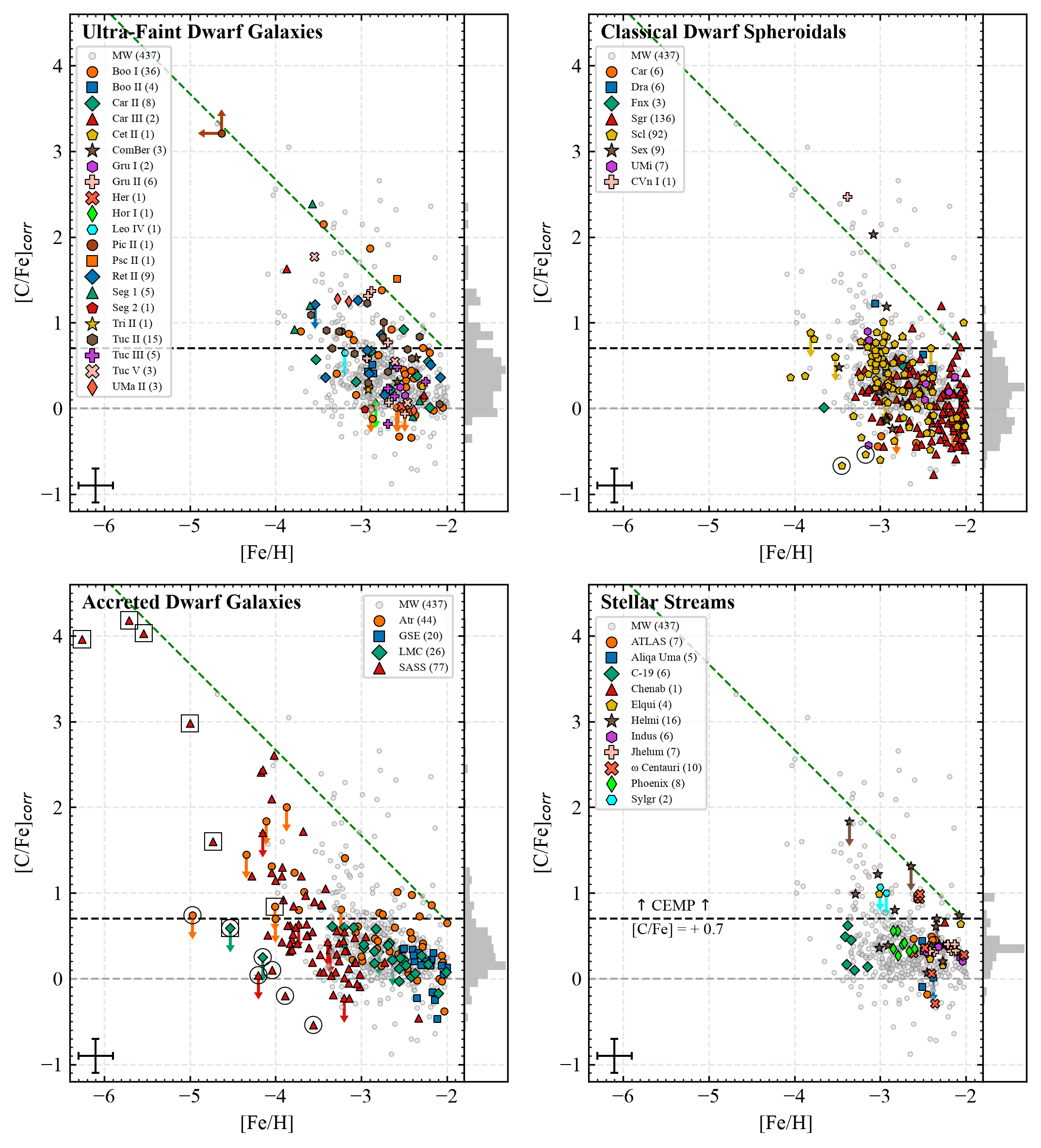}
    \caption{\small
        \feh\ vs \cfe\ for stars in each of the four categories of systems. 
        The legend provides information on the individual systems in each category. 
        Milky Way stars are shown as light gray points for comparison. 
        The dashed dark gray line at \cfe\ $=0.0$ represents the solar carbon abundance. 
        The black dashed line at \cfe\ $=0.7$ represents the CEMP threshold, while the gray dashed line at \cfe\ $=0.0$ depicts the solar carbon abundance ratio.
        The green dashed line at $A(C) \simeq 7.1$ marks the approximate separation of CEMP-$s$/CEMP-$i$ stars and CEMP-$no$ stars \citep{Yoon2016a} used for cleaning the sample.
        Typical error bars of $\pm 0.2$\,dex are shown at the bottom left of each panel.
        The histograms showcase the \cfe\ distribution of the respective panels.
    }
    \label{fig:feh_cfe}
\end{figure*}

% carbon correction
Due to internal mixing caused by convection during late stages of stellar evolution, observed carbon abundances in the stellar atmosphere are reduced from those present in the natal gas cloud.
Thus, the observed carbon abundances in our sample require a correction in order to establish their natal carbon abundances. 
Following \citet{Placco2014c}, we applied these corrections to all the carbon abundances (consistently using the solar abundances from \citealt{Asplund2009a}) depending on the evolutionary status of each star.
These evolutionary-corrections increased the carbon abundances of the individual stars from anywhere between 0.0 to 0.8\,dex, with values shown in Table~\ref{tab:abunds1}.
Going forward, when we refer to any carbon abundances, we mean the evolutionarily-corrected values.

For stars with upper limit measurements on \cfe, we note that any correction to the carbon would depend on the true, unknown carbon abundance.
In the absence thereof and to mitigate an over-correction (to a reasonable extent), we chose to calculate the corrections based on the stars' upper limit value minus $0.3\,\rm{dex}$.
The $0.3\,{\rm dex}$ adopted follows a conservative estimate of the carbon abundance uncertainty.
This reduces the corrections by $\sim 0.05 - 0.1\,{\rm dex}$.
The corrections were then applied to the unmodified upper limits values.
To note, this treatment of the evolutionary-corrections to the \cfe\ upper limits has very little effect on the CEMP fractions discussed in Section~\ref{sec:cempfractions}.

% removing extrinsic carbon enrichment
Based on the \cfe\ $\ge 0.7$ criterion for carbon enhancement, we established an initial group of CEMP stars with corrected carbon values.
This subsample was then inspected for stars whose carbon abundances may have originated from extrinsic enrichment (rather than from a previous stellar generation), such as those within binary systems where carbon-rich material is transferred between companions. 
This includes the \sproc\ metal-poor stars, as well as \iproc\ stars (previously known as $r/s$ stars), typically identified by their barium and europium abundances: $\rm{[Ba/Eu]} > 0.5$ and $\rm{[Ba/Fe]} > 0.6$ \citep{Beers2005b, Placco2014c, Frebel2018b}.
For stars without europium abundances available, we instead identified the \sproc\ nature in CEMP stars using their strontium and barium abundances, requiring \cfe\ $\ge 0.7$, $\rm{[Sr/Ba]} < -0.5$, and $\rm{[Ba/Fe]} > 0.6$ \citep{Aoki2013b, Spite2014b, Placco2014c}.
CEMP-$s$ stars may also be identified by their extreme carbon abundances, following the separation of CEMP-$s$/CEMP-$i$ stars from CEMP-$no$ stars at $A({\rm C}) = 7.1$ \citep{Bonifacio2015a, Yoon2016a}.

After identifying a sub-sample of CEMP-$s$/CEMP-$i$ candidates via any of the three criteria above, we further inspected their abundance patterns and literature before classifying these stars as CEMP-$s$/CEMP-$i$ and removing them from our initial sample of CEMP stars.
The stars remaining above the green-dashed line representing the \citet{Yoon2016a} criterion in Figure~\ref{fig:feh_cfe} were classified as CEMP-$no$ stars (\cfe\ $\ge 0.7$ and $\rm{[Ba/Fe]} \le 0.0$) based on either their measured abundances or existing literature classifications, or they lacked the neutron-capture abundances necessary for proper classification.
These stars remain in our sample.

% final sample
All in all, this preparation resulted in a sample size of 608 stars between all the dwarf galaxies, SASS stars, and stellar streams considered in this study. 
This includes 109 stars in twenty-one UFDs, 260 stars in eight CDWs, 90 stars in three ADWs, 77 SASS stars, and 72 stars in eleven stellar streams.
Included in this sample are 31 CEMP stars in the ultra-faint dwarf galaxies, 25 in the classical dwarf spheroidal galaxies, 35 in the accreted dwarf galaxies (24 of which are SASS stars), and 9 in the stellar streams.
Details on which system these stars belong to and total sample sizes can be seen in Table~\ref{tab:sysproperties}.
In Table~\ref{tab:abunds1}, we present selected stellar abundances for all Milky Way, dwarf galaxy, and stream stars considered in this work, including iron, carbon, silicon, barium, strontium, and europium abundances (when available) along with their associated literature reference.

% milky way (for comparison)
For the comparison with the Milky Way metal-poor halo stars, we use the collection of high-resolution stellar chemical abundances from \citet{Placco2014c}. 
Here, we already note a difference in the Milky Way CEMP star fractions used in our comparison relative to those in \citet{Placco2014c} due to the updates to the sample made as part of this paper, as described below.

From the Placco et al. sample, we removed all stars classified as members of the Atari Disk \citep{Mardini2022a} -- including SDSS~J102915.15+172927.9 \citep{Caffau2011d} -- while retaining only a single entry for stars with multiple carbon abundance measurements to avoid accidental duplication.
Similarly, we removed an additional 43 SASS stars from the Milky Way sample, identified by the selection criteria of $\rm{[Sr/H]} \le -4.5$ and $\rm{[Ba/H]} \le -4.0$, by supplementing the Placco et al. sample with abundances from JINAbase.
We note that the removal of these known systems and sub-structures in the Milky Way halo sample does not necessarily mean the remaining stars formed in-situ; the majority of the halo is likely composed of accreted stars whose progenitor systems have yet to be identified \citep{Deason2016a, Deason2023a}.
The same carbon corrections and selection procedure were then applied to remove any stars likely to be CEMP-$s$ and CEMP-$i$/CEMP-$r/s$ stars. 
With this done, there are 79 CEMP stars with \cfe\ $\ge 0.7$ after carbon correction, and a total of 437 stars remaining in our Milky Way sample.
The Milky Way stellar abundances and associated references can also be found in Table~\ref{tab:abunds1}.  

% the [C/Fe] figure
The corrected carbon abundances of the stars in the various dwarf galaxies, streams, and the Milky Way are shown in Figure~\ref{fig:feh_cfe}, grouped by each of the four categories of systems.
We have chosen to use the mean uncertainty of the carbon abundance in our sample ($\pm0.2\,\rm{dex}$) as a typical error on all measured carbon abundances.
The differences between these categories can clearly be seen in the histograms along the y-axes, showing the \cfe\ distribution (excluding the lower and upper limits).
For example, the extended high \cfe\ tails found in among ultra-faint dwarf galaxies and SASS stars are not found in the other categories.

\input{tables/tab2_abundances1}
\label{tab:abunds1}

\input{tables/tab3_abundances2}
\label{tab:abunds2}

%=================================================
\subsection{Limitations and considerations regarding the carbon abundance compilation}

This compilation of literature stellar abundances is heterogeneous by design, comprising analyses carried out by many individual investigators and collaborations using different methodologies.
While we do not attempt to catalog every methodological difference among the $\sim$115 literature sources used (see \citealt{Arentsen2022a} for a similar discussion), we here briefly comment on the main potential sources of inhomogeneity in our carbon abundance compilation.
This includes target selection procedures, determinations of stellar atmospheric parameters, choices of 1D/3D model atmospheres, and the assumption of LTE (versus correcting for NLTE).

% Stellar atmospheric parameters
For the atmospheric stellar parameters, there are large difference between photometric and spectroscopic temperatures, especially for cool RGB stars \citep{Frebel2013b}.
These differences impact the \ch\ abundances; however, \cfe\ values should be less affected overall due to being an abundance ratio with \feh.

% LTE/NLTE and 1D/3D
Regarding choices about LTE/NLTE and the use of 1D/3D model atmospheres, we adopted the published 1D LTE \feh\ abundances to preserve internal consistency across the literature.
A few 3D studies are available for carbon abundances in the literature, and estimates for 3D and NLTE corrections for carbon abundances in metal-poor stars were developed in \citet{Norris2019a}, to which we refer the interested reader.
However, given the size and heterogeneity of our sample, we do not attempt to apply any such corrections here.

% Photometric Selection Effects
The techniques used for metal-poor candidate selection that leads to high-resolution spectroscopic follow-up, and hence sample completeness, are important to consider as well.
A long-established method involves obtaining \feh\ estimates by measuring the Ca\,K (CaK) line at 3933.7\,\AA\ in low- or medium-resolution spectra \citep{Beers1999a}, without regard to any information of the carbon abundance.
Prior to $\sim$2015, the high-resolution abundances reported in the majority of papers, including our Milky Way halo sample \citep{Placco2014c}, utilized in our data compilation were ultimately based on this technique.

Starting with \citet{Keller2014a} and \citet{Jacobson2015a}, narrow-band photometry began to be more commonly used to select metal-poor candidates.
The narrow photometric bandpass filters initially used captures the CaK region, in addition to the neighboring CN molecular band near $3883\,$\AA.
For carbon- and nitrogen-rich stars, the CN feature can contribute to the total absorption within the filter, making a star appear photometrically more metal-rich than what is truly is.
In individual cases, a carbon-rich, iron-poor star could therefore be inadvertently left out from any high-resolution spectroscopic follow-up; however, the extent of any such bias would depend on the metallicity and effective temperature of the star in question and may not have significant consequences in most cases \citep{Chiti2020a}.
To alleviate this principal issue, the Pristine survey \citep{Starkenburg2017b} has used a modified narrow-band filter designed to more cleanly isolate the CaK feature from CN contamination.
As a result, the bias against particular metal-poor stars has become substantially less pronounced.

Specifically with regards to our sample, this bias could arise within only a few selected studies, as their samples were selected based on the broader CaK filter: \citet{Ou2025a} about Sagittarius (35 stars) and \citet{Chiti2018b} about Tucana II (7 stars).
This could affect the CEMP fraction in the CDWs and UFDs (discussed in detail below).

Regarding UFDs, we note that essentially all stars bright enough for detailed spectroscopic follow-up are usually observed, regardless of their metallicity, limiting the potential for any bias in these systems.
Besides that, if a carbon bias were present, then any reported CEMP fractions for the UFDs would be even higher. This would further distinguish our UFD results from those for the CDWs (see also below).
For the CDWs, the CEMP fraction may also be underestimated due to inclusion of the \citet{Ou2025a} study; however, in comparison to the other stars observed in Sagittarius, the CEMP fraction of the \citet{Ou2025a} stars is not significantly different.

%%%%%%%%%%%%%%%%%%%%%%%%%%%%%%%%%%%%%%%%%%%%%%%%%%%%
\section{Fraction of CEMP stars across different dwarf galaxies} \label{sec:cempfractions}

We calculate the CEMP star fractions based on the corrected \cfe\ abundances for the different dwarf galaxies and streams in order to assess potential differences in the carbon production across early galaxies.
The CEMP star fraction is defined by the number of CEMP stars over the total number of stars in our finalized sample ($N_{\rm CEMP} / N_{\rm total}$) for a given metallicity range.
Stars with upper limits of \cfe\ $\le 0.9$ are included in the total number of stars ($N_{\rm total}$) as they have a (relatively) low likelihood of being classified as CEMP stars in the future.\footnote{There are a total of eight stars with \cfe\ upper limits affected by this method of counting, including four Atari stars, three Sculptor stars, and one Helmi star. Their inclusion only changes the MC CEMP fraction of the individual systems by $\lesssim 2\%$, well within uncertainty.}
However, we cannot classify any stars with high carbon upper limits as bona-fide CEMP stars, as their true carbon abundance is unknown; thus, we exclude them from being counted in $N_{\rm CEMP}$ and $N_{\rm total}$.
When adopting the \cfe\ $=1.0$ criterion for CEMP star classification, we similarly include stars with upper limits of \cfe\ $\le 1.2$.
Because of this, the $N_{\rm total}$ value of the CEMP star fraction is occasionally reduced for a given CEMP classification threshold, relative to the actual total number of upper limits and measurements available in our sample.
We also note that there are two stars with upper limits in \feh\ in our sample that are included in $N_{\rm CEMP}$ and $N_{\rm total}$: a star in Pictor\,II (Pic~II$-$503) with \cfe\ $\ge 3.2$ \citep{Chiti2025a} and a SASS star in the halo (SMSS~J031300.36$-$670839.3) with \cfe\ $\ge 5.0$ \citep{Keller2014a}.
We don't show the latter star in Figure~\ref{fig:feh_cfe} and Figure~\ref{fig:feh_dtrans} given its extreme abundances.

\begin{figure}[!htb]
    \centering
    \includegraphics[width=3.3in]{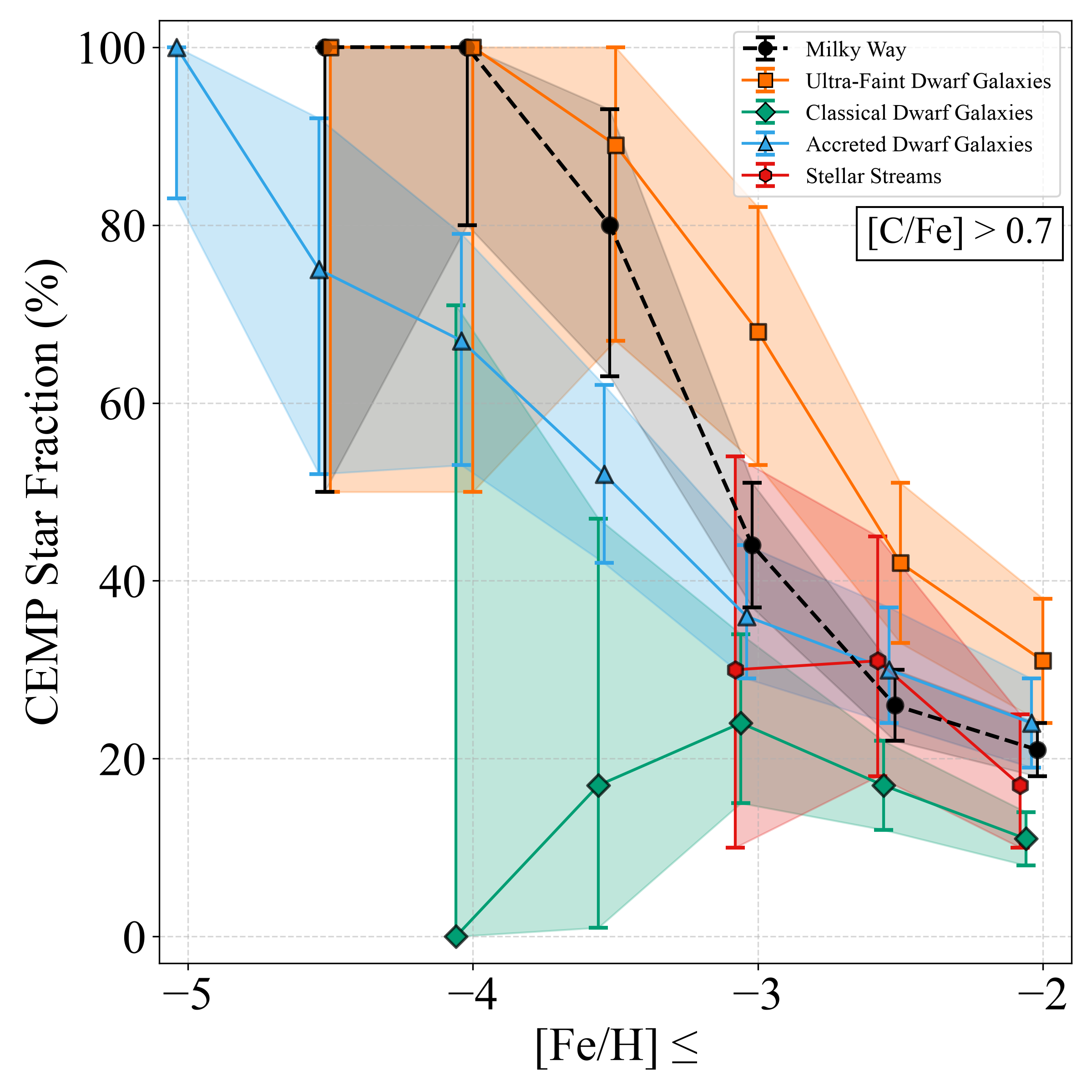}
    \caption{\small
        \feh\ vs. MC CEMP star fractions for stars in each category of systems.
        CEMP fractions are depicted for UFDs (orange line), CDW (green line), SSs (red line), and ADWs (SASS stars: blue dashed line, accreted dwarf galaxy stars: dotted blue line, and cumulative accreted stars: solid blue line). 
        Milky Way stars are shown in black for comparison.
    }
    \label{fig:cempfractions}
\end{figure}

Table~\ref{tab:cempfractions} presents the ``Observed'' and ``Monte Carlo'' (MC) CEMP star fractions for multiple metallicity bins and two carbon-enhancement thresholds: \cfe\ $\le 0.7$ and \cfe\ $\le 1.0$ (to assess robustness).
The ``Observed'' values utilize the at-face corrected carbon abundances presented in the Table~\ref{tab:abunds1}.
In comparison, we also derive the ``Monte Carlo'' values to better account for the uncertainties on the CEMP fractions and the overall \cfe\ distributions.
Throughout the rest of this paper, we generally use the MC CEMP fractions when comparing systems and categories.

Uncertainties stem from two contributions: the uncertainty of whether a star can be classified as a CEMP star and the uncertainty that comes from the overall sample sizes.
To address the classification uncertainty, we resample each carbon abundance measurement and upper limit over a normal distribution with $\sigma = \pm 0.2$\,dex in a Monte Carlo simulation, and calculate the CEMP star fraction and associated spread based on the resulting distribution.
To address the sample size uncertainty, we use the Wilson score interval for binomial statistics, which both handles small samples effectively and produces asymmetric uncertainties near the 0\% and 100\% bounds.
These two sources of uncertainty were combined linearly as an approximation of the total uncertainty, as they are heavily correlated, and re-bounded from 0\% to 100\%.
The primary source of the uncertainty originates from the sample sizes, and as a result, the uncertainty is larger at lower metallicities where fewer carbon abundances are available.
In Figure~\ref{fig:cempfractions}, the resulting MC CEMP fractions and uncertainty bands are shown.
The individual CEMP star fractions of each system can be found in Table~\ref{tab:sysproperties}. 
Additional details on selected systems CEMP fractions are discussed in Appendices~\ref{sec:appendix_UFD}-\ref{sec:appendix_SS}.

\medskip
In the following, we provide an overview of the CEMP fractions in each category, discuss the available observations and their contributions to these fractions, and compare the fractions across categories and with that of the Milky Way halo.

%=================================================
\subsection{Ultra-faint Dwarf Galaxies} \label{sec:UFD}

Among the 21 ultra-faint dwarf (UFDs), which together contain 109 stars in our sample, 11 systems host a total of 31 observed CEMP stars, while many additional stars reside just below the \cfe\ $=0.7$ threshold.
This corresponds to an overall MC CEMP fraction of $31\%_{-07}^{+07}$ (34 CEMP stars of 107) for stars with \feh\ $<-2.0$.
Generally, the UFD CEMP fraction is significant across all low metallicities and is consistently greater than or equal to that of the Milky Way.
For example, if we consider the stars with \feh\ $\le -3.0$, the UFD CEMP fraction reaches $68\%_{-15}^{+14}$ whereas the Milky Way halo CEMP fraction only reaches $44\%_{-07}^{+07}$, as seen in Figure~\ref{fig:cempfractions} and Table~\ref{tab:cempfractions}.

However, 16 of the 21 UFDs have five or fewer stars with any carbon abundance measurements (13 of the 21 UFDs have three or fewer stars). 
Given the sparsity of data (e.g., Cetus\,II, Hercules, Horologium\,I, Leo\,IV, Pictor\,II, Pisces\,II, Segue 2, and Triangulum\,II each only have one star observed), CEMP stars may well be present in these 16 systems as well.
This notion is principally supported by the four of these 16 systems that possess CEMP stars: Carina\,III (1 CEMP star of 2) has a CEMP fraction of $50\%_{-32}^{+32}$, Segue\,1 (3 of 5) has $60\%_{-29}^{+25}$, Tucana\,V (1 of 3) has $33\%_{-33}^{+42}$, and Ursa Major\,II (2 of 3) has $67\%_{-20}^{+22}$.
When considering only the five systems with at least six stars with carbon measurements (Bootes\,I, Carina\,II, Grus\,II, Reticulum\,II, and Tucana\,II), 23 stars of the 72 are classified as CEMP stars, yielding a CEMP fraction of $32\%_{-09}^{+09}$ for \feh\ $< -2.0$ -- a value equivalent to the result obtained for the full UFD sample.

%=================================================
\subsection{Classical Dwarf Spheroidal Galaxies} \label{sec:CDW}

For the eight classical dwarf spheroidal galaxies (dSph; CDW), which together contain 260 stars in our sample, six of the systems host a total of 25 observed CEMP stars and an overall MC CEMP fraction of $11\%_{-03}^{+03}$ (29 CEMP stars of 258) for \feh\ $<-2.0$.
Neither Carina (6 stars) nor Fornax (3 stars) possess CEMP stars in their respective samples.
In contrast to the other categories of systems, the CDWs have the lowest overall CEMP fraction across all metallicities.
For instance, there are no CEMP stars are found with \feh\ $\le -4.0$ ($0\%_{-00}^{+71}$), and only one found with \feh\ $\le -3.5$ ($17\%_{-16}^{+30}$). 
The CEMP fraction is highest ($24\%_{-09}^{+10}$) when considering only stars with \feh\ $\le -3.0$, but this still remains below that of the Milky Way halo and the other categories in the same metallicity regime.

Sagittarius and Sculptor together are the primary contributors to the overall CDW sample, containing $\sim 90\%$ (228 stars) of all the CDW stars in our final sample alone.
Their combined CEMP fraction is  $10\%_{-03}^{+04}$ (22 CEMP stars of 226) for stars with \feh\ $\le -2.0$, equivalent to the overall CDW CEMP fraction.
The remaining CEMP stars are found in Canes Venatici I (1 CEMP star out of 1), Draco (1 CEMP star out of 6), Sextans (2 out of 9), and Ursa Minor (2 out of 7). 
No individual system has a CEMP fraction greater than $\sim 30\%$ (excluding Canes Venatici I with only one carbon abundance measurement), and carbon enhancement is generally modest: carbon abundances rarely exceed \cfe\ $=0.9$, and only five stars have \cfe\ $\ge 1.0$. 
Overall, these eight CDW systems show little evidence for strong carbon enhancement among the metal-poor stellar population.

%=================================================
\subsection{Accreted Dwarf Galaxies} \label{sec:ADW}

For the accreted dwarf galaxies (ADWs), we considered systems composed of the merger debris of disrupted dwarf galaxies now completely present in the local halo, bulge, and thick disk of the Milky Way.
This includes stars associated with the Atari Disk (Atr; \citealt{Mardini2022a, Mardini2024a}), Gaia-Sausage/Enceladus (GSE; \citealt{Haywood2018a, Helmi2018a, Belokurov2018b, Naidu2020c, Kruijssen2020c, Limberg2022a}), and Large Magellanic Cloud (LMC; \citealt{Chiti2024a, Ji2026a}).

For these three systems, which together contain 90 stars in our sample, there are only 11 observed CEMP stars and an overall MC CEMP fraction of $15\%_{-06}^{+07}$ (13 CEMP stars of 86) for \feh\ $<-2.0$.
Interestingly, all 11 of these observed CEMP stars are found in the Atari sample. 
The additional CEMP stars used our MC CEMP fraction originate statistically from Atari and LMC stars near the carbon-enhancement threshold; however, we note that none of the LMC stars are actually observed with \cfe\ $\ge 0.7$ (shown in  Figure~\ref{fig:feh_cfe}). 
The GSE currently has no known CEMP stars, but that is likely due to a lack of known extremely metal-poor stars.
Still, overall, these three systems show a trend similar to that of the Milky Way, namely an increasing CEMP fraction towards lower metallicities for $-3.5 \le$ \feh\ $\le -2.0$.

In addition to these three systems, we separately consider the SASS stars within the ADW category. 
There are 24 observed CEMP stars out of the 77 available SASS stars, yielding a MC CEMP fraction of $36\%_{-08}^{+08}$ (27 CEMP stars of 76) for \feh\ $<-2.0$.
This sample exhibits a strong metallicity dependence, with the CEMP fraction increasing significantly below \feh\ $\le -3.5$. 
At the lowest metallicities, SASS stars are the primary or sole contributors to the overall ADW CEMP fraction, driving it to $67\%_{-14}^{+12}$ for \feh\ $\le-4.0$, which continually increases to $75\%_{-23}^{+17}$ for \feh\ $\le-4.5$ and $100\%_{-17}^{+00}$ for \feh\ $\le -5.0$.
We further note that the SASS stars have the highest \cfe\ abundances relative to any in any of the other systems, with the largest \cfe\ abundances occurring at the lowest metallicities. 

If one were to re-calculate the Milky Way halo CEMP fraction without the removal of the ADW stars, the MW halo CEMP fractions more closely aligns with the ADW CEMP fractions due to the increased number of stars with \feh\ $\le-3.5$ where the SASS population is dominate.

\input{tables/tab4_cempfractions}
\label{tab:cempfractions}

%=================================================
\subsection{Stellar Streams} \label{sec:SS}

Among the eleven stellar streams (SS) considered here, comprising a total of 72 metal-poor stars with $-3.5 \le$ \feh\ $\le -2.0$, the available carbon abundance measurements remain sparse. 
These systems include remnants of four disrupted globular clusters (ATLAS, Aliqa Uma, C-19, and Phoenix), five dwarf galaxy progenitors (Chenab, Elqui, Indus, Jhelum, and $\omega$~Centauri), and two systems of uncertain origin (Sylgr and Helmi). 
Much of the high-resolution spectroscopic data has been obtained through the Southern Stellar Stream Spectroscopic Survey (${\rm S}^{5}$) \citep{Ji2020b}, with additional measurements compiled from \citet{Gull2021a, Roederer2010a, Martin2022a}. 
Each system contains at most eight metal-poor stars with detailed abundances, highlighting the general difficulty of finding stellar streams at low metallicity.
Only two systems -- Helmi stream (16 stars) and $\omega$~Centauri (10 stars) -- have sample sizes exceeding ten stars with \feh\ $<-2.0$.

CEMP stars are observed in only three of these systems: the $\omega$~Centauri stream (4 CEMP stars out of 10), the Helmi stream (4 out of 14, excluding two with high upper limits), and Elqui (1 out of 4).
No CEMP stars have been observed in the remaining eight systems.
This absence is not necessarily unexpected though given the limited number of stars observed per system.
This is particularly true for the Chenab and Sylgr streams, where each is represented by less than four metal-poor stars.
When considered collectively and at face value, the stellar stream sample exhibits an overall MC CEMP star fraction of $17\%_{-07}^{+08}$ (12 CEMP stars out of 68) for \feh\ $\le -2.0$.
This is comparable to the $21\%_{-03}^{+03}$ of the Milky Way halo in the same metallicity regime.
Not surprisingly, this suggest that disrupted satellites and clusters contribute meaningfully to the CEMP population of the stellar halo.

%===================================================
\medskip
In the following Sections, we broadly examine and explore several complementary diagnostics based on the CEMP fractions and the \ch\ abundances to construct an observationally motivated framework of early galaxy formation and associated formation channels for the different categories of surviving systems, as well as the halo of the Milky Way.

%%%%%%%%%%%%%%%%%%%%%%%%%%%%%%%%%%%%%%%%%%%%%%%%%%%%
\section{CEMP star fractions reflecting early galaxy formation channels} \label{sec:galaxy_formation}

In Figure~\ref{fig:feh_cfe}, we presented \cfe\ abundances as a function of metallicity for various systems. 
Motivated by the tendency of the more massive dwarf galaxies  to have a lower occurrence rate of CEMP stars compared to the lower-mass UFDs, we now examine how the CEMP fractions among the categories of systems (UFDs, CDWs, ADWs, and SSs) are reflective of different early galaxy formation pathways and channels.

%=================================================
\subsection{Existence of the Magnitude--CEMP Fraction (MCF) relation at low metallicity} \label{sec:mcf_relation}

\begin{figure}[!htb]
    \centering
    \includegraphics[width=3.3in]{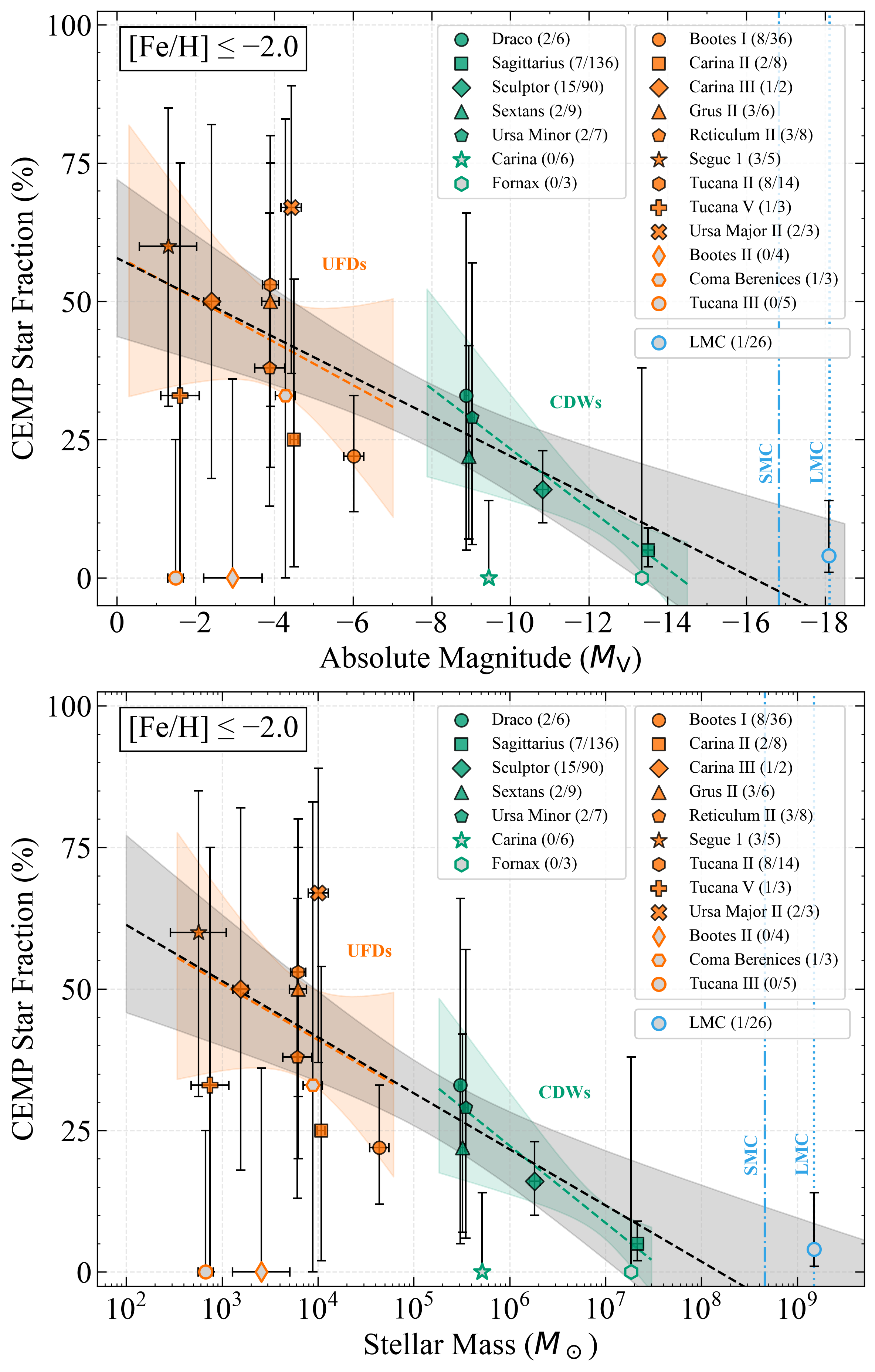}
    \caption{\small
        Magnitude--CEMP Fraction (MCF) relation for dwarf galaxy systems containing CEMP stars with \feh\ $\le -2.0$.
        The top panel shows the relation between the absolute $V$-band magnitudes ($M_{\rm V}$) and the MC CEMP star fractions of the different systems.
        The bottom panel shows the equivalent relation using systems' stellar mass.
        The MC CEMP fraction for each system is notated in the legend.
        Data points and fitted trends are color-coded by system type: CDW are shown in green, UFDs in orange, and ADW in blue.
        The black line shows the overall trend obtained when fitting the CDW and UFD samples together.
        The systems with gray data points not included in the fitted trends.
        The blue dotted and dot-dashed lines indicate the LMC and SMC values, respectively.
    }
    \label{fig:mcf_relation}
\end{figure}

% fit parameters
We first investigate how the MC CEMP star fractions vary with the host systems properties.
For stars with \feh\ $\le -2.0$, Figure~\ref{fig:mcf_relation} reveals a new Magnitude--CEMP Fraction (MCF) relation in terms of each system's absolute $V$-band magnitude ($M_{\rm V}$; top panel) and stellar mass ($M_{\rm stellar}$; bottom panel), where the latter is estimated by adopting a stellar mass-to-light ratio of one for CDWs and the LMC, and two for UFDs \citep{Simon2019a, Martin2008a}.
We note that we cautiously only considered the systems that contain at least two metal-poor stars with carbon abundance measurements and contain at least one CEMP star for the fits.
We also show three UFDs and two CDWs with gray points in Figure~\ref{fig:mcf_relation} that are not included in the fits.
These systems possess at least three stars with carbon abundance measurements, none of which have been observed as carbon-enhanced.

The MCF relation is characterized by a significant overall trend of the CEMP star fraction increasing toward fainter, less massive systems.
This is illustrated by the linear fit for the combined CDW and UFD systems, with best-fit parameters and standard deviations of $\left[ m,\ b \right] = [3.6\pm1.6,\ 58\pm14]$ (black dashed line).
When applying a linear fit to the two categories separately, we find the best-fit parameters of $\left[ m,\ b \right] = [5.4\pm3.6,\ 78\pm45]$ for CDWs (green line) and $\left[ m,\ b \right] = [3.9\pm6.1,\ 58\pm26]$ for UFDs (orange line).
Translating the magnitudes into host stellar masses, the same behavior is analogously seen in the stellar-mass plot (bottom panel of Figure~\ref{fig:mcf_relation}).
In this case, the best-fit parameters and standard deviations are $\left[ m,\ b \right] = [-13.6\pm9.0,\ 104\pm63]$ for CDWs, $\left[ m,\ b \right] = [-9.9\pm15.2,\ 81\pm60]$ for UFDs, and $\left[ m,\ b \right] = [-9.9\pm4.5,\ 81\pm24]$ for the combined trend.
Note that uncertainties on the magnitudes and stellar masses were not used in the fitting process.
     
% uncertainties of data and fit
The uncertainties on the MC CEMP fractions for individual systems are calculated as described in Section~\ref{sec:cempfractions}, with the dominant contribution arising from small sample sizes.
To estimate the uncertainty on the fitted relations, we propagate these uncertainties through bootstrap resampling, weighting each system by its corresponding CEMP fraction uncertainty.
In the MCF relation, the resulting uncertainty bands span $\sim 15\%$-$45\%$ for the UFDs (orange), $\sim 5\%$--$30\%$ for the CDWs (green), and $\sim 10\%$--$30\%$ for the combined trend (black).
The uncertainty ranges in the stellar mass plot are comparable.
Increasing sample sizes in the future will greatly reduce these uncertainties.

% prediction of other system's CEMP fractions
We use the MCF relation to predict the CEMP fractions of systems that currently lack CEMP stars with \feh\ $\le-2.0$ to be included in the fits of Figure~\ref{fig:mcf_relation}.
For the UFD systems at the low-mass end, Coma Berenices (3 stars; $M_{\rm V} = -4.28_{-0.25}^{+0.25}$), Bootes II (4 stars; $M_{\rm V} = -2.94_{-0.75}^{+0.74}$), and Tucana III (5 stars; $M_{\rm V} = -1.49_{-0.20}^{+0.20}$) all lack observed CEMP stars.
The carbon abundances measured in these systems all have \cfe\ values well within the bulk of the observed distribution, ranging from $-0.2$ to $0.6$.
Yet, following the fitted trends, we expect to find many additional CEMP stars at these magnitudes, with each individual systems hosting CEMP fractions of $\sim 45-50 \pm 15 \%$.
For the CDW systems at the high-mass end, Carina and Fornax similarly lack confirmed CEMP stars.
In Carina, all six available carbon measurements fall below \cfe\ $< 0$.
With a $V$-magnitude of $M_{\rm V} = -9.45_{-0.05}^{+0.05}$, we expect Carina to host a CEMP fraction of $\sim 25 \pm 10 \%$ for stars with \feh\ $\le -2.0$.
Fornax, being more luminous and massive ($M_{\rm V} = -13.34_{-0.14}^{+0.14}$), is expected to host an even smaller CEMP fraction of order $\lesssim 5-15\%$.
Of three currently available Fornax stars with carbon measurements, none are considered carbon-enhanced.
In any of these systems, larger carbon-abundance samples from future observations should bring these systems' observed CEMP fractions closer to their predicted values by providing at least a few CEMP stars.

We further predict CEMP fractions for the Magellanic Clouds.
Since carbon abundances are not yet available for the SMC, and the LMC is treated separately as an accreted dwarf system \citep{Besla2007b}, neither galaxy was included in the fits. 
The magnitudes and stellar masses of both the SMC ($M_{\rm V} = -16.8_{-0.1}^{+0.1}$, $M_{\rm stellar} = 4.6_{-0.4}^{+0.4} \times 10^{8}\,M_{\odot}$) and LMC ($M_{\rm V} = -18.1_{-0.1}^{+0.1}$, $M_{\rm stellar} = 1.5_{-0.1}^{+0.1} \times 10^{9}\,M_{\odot}$) are indicated by vertical blue lines in Figure~\ref{fig:mcf_relation}.
The MCF relation predicts that only a small CEMP fraction should be present in these massive systems.
At present, the measured CEMP star fraction of the LMC (and associated uncertainty range) agrees very well with the extrapolated relation.
This strongly suggests that the low-metallicity MCF relation may principally be valid for satellite systems with detected CEMP stars over a very wide range of host stellar masses, spanning roughly from $10^{2}$ to $10^{9}\,M_{\odot}$.

%=================================================
\subsection{``Faint'' supernovae and hypernovae drive early galaxy formation  channels} \label{sec:supernovae}

We now consider the nucleosynthetic origins of carbon in the earliest generations of stars to learn about galaxy formation channels. 
``Faint'' supernovae, characterized by extensive fallback, are thought to eject large amounts of carbon while producing relatively smaller amounts of iron. 
The putative existence of these supernovae has been used to explain the very high \cfe\ ratios of the most iron-poor stars in the Milky Way halo \citep{Umeda2003a} and, collectively, the associated high CEMP fraction at low metallicity \citep{Cooke2014b}.
Aspherical supernovae and rapidly rotating massive Pop\,III stars are also thought to produce high \cfe\ yields, although we do not consider them here as primary sources of the highest carbon abundances \citep{Tominaga2009a, Ezzeddine2019a}.
On the other hand, stars with lower, more ``normal'' carbon abundances closer to the solar ratio are likely the result of enrichment from early more luminous, energetic supernovae (``hypernovae''; \citealt{Umeda2005a, Tominaga2007b, Heger2010a}) that do not undergo any fallback process.
This raises the question of how early nucleosynthesis, together with mixing, dilution, and gas retention processes, can lead to signatures that are preserved in both CEMP stars and the broader stellar population with more typical carbon abundances of $-0.1 \lesssim$ \cfe\ $\lesssim 0.6$ across different systems.

% UFD and CDW early chemical enrichment
The high CEMP fraction found for UFDs strongly suggest that these systems formed in environments that experienced a relatively large contribution from faint supernovae compared to more energetic core-collapse events \citep{Cooke2014b, Chiti2026a}.
This is in line with other suggestions that UFDs may be the original hosts of the most metal-poor halo stars \citep{Frebel2014a, Frebel2015a}.
In contrast, the more massive systems, such as the CDWs, primarily host stars with $-0.5 \lesssim$ \cfe\ $\lesssim 0.5$, although a small fraction of CEMP stars are still present (see Table~\ref{tab:cempfractions}).
These predominantly ``normal'' \cfe\ stellar populations suggest that the early assembly and enrichment histories of CDWs fundamentally differed from those of the UFDs.

Several formation pathways for CDWs could, in principle, explain this important difference:
\begin{enumerate}
    \item CDWs exclusively formed through the mergers of a large number of smaller UFD-like systems that all did not experience a significant number of early faint supernovae.
    However, this scenario appears unlikely given the high CEMP star fractions observed in the majority present-day UFDs, as discussed in Section~\ref{sec:cempfractions}. 
\vspace{-0.2cm}
    \item CDWs were assembled from UFD-like systems that hosted a substantial, nearly coeval mix of faint supernovae and hypernovae, such that the carbon-enhancement signatures from the faint supernovae were rapidly diluted before later generations of lower mass stars formed.
    However, this scenario also appears unlikely as it would require a significant number of hypernovae to occur promptly at early times to erase the strong CEMP signature found from the UFDs and among halo stars.
\vspace{-0.2cm}
    \item CDWs did not assemble predominantly from smaller UFD-like systems, but instead formed directly from more massive gas reservoirs and the enrichment was dominated by energetic hypernovae rather than faint supernovae.
    A limited number of UFD-like systems may have been accreted at later times, which would then account for the small but non-zero CEMP fraction present in CDWs.
\end{enumerate}

Overall, the observed carbon abundances most strongly favor the third scenario.
Rather than being scaled-up analogues of early UFD-like systems assembled through a simple hierarchical process, surviving CDWs appear to have formed through a distinct early enrichment pathway in which faint supernovae played only a minor role.
As such, a substantial fraction of their stellar mass must have originated in more massive star-forming systems with intrinsically lower CEMP fractions, supplemented by the accretion of less-massive UFD-like systems with higher CEMP fractions.
This UFD-like contribution becomes progressively less dominant with increasing galaxy mass, indicating that early systems across different mass regimes acted as chemically distinct building blocks in hierarchical galaxy formation, each reflecting a different formation pathway.

To test this formation pathway, we consider the accreted dwarf systems and stellar streams and assess whether our interpretation persists beyond today's surviving dwarf satellites.

% accreted dwarf galaxies 
For instance, the two more massive ADW systems (LMC and GSE) currently contain no observed CEMP stars.
Consequently, they could have plausibly assembled from CDW-like systems, as expected within hierarchical assembly.
On the contrary, the SASS star sample includes some of the most iron-poor and carbon-enhanced stars found to date.
Each SASS star, by design, represents an early accreted primitive host system formed from gas chemically enriched by only a small number of early supernovae, as indicated by their paucity of neutron-capture abundances.
Many of these early supernovae must have been faint, given the large CEMP fraction found across the SASS sample.
As a result, it appears that the associated early host systems share key characteristics with those inferred for the UFDs.
Interestingly, the Atari Disk, thought to be an ancient system accreted very early in the Milky Way’s assembly history, displays a mixture of ``normal'' and enhanced \cfe\ values.
As such, the abundance signature of the original host system may have been the result of either the merger of multiple UFD-like progenitors (Scenario 1), or it originated from UFD-like substructures embedded within a larger gas reservoir (similar to Scenario 3) without the erasure of the faint-supernovae signature \citep{Mardini2022a}.

% stellar streams
For the stellar stream stars likely associated with disrupted dwarf galaxies (Chenab, Elqui, Indus, Jhelum, $\omega$~Centauri, and Helmi), most exhibit ``normal'' carbon abundance levels with roughly a fifth showing carbon enhancement ($24\%_{-10}^{+11}$, 10 CEMP stars out of 42).
Although the pre-disruption masses of these systems are difficult to constrain, this level of carbon enhancement is broadly consistent with that observed in the UFDs over a similar metallicity range.
We thus speculate that the former hosts may not have been overly massive.
On the other hand, the stream stars originating from globular clusters (ATLAS, Aliqa Uma, C-19, and Phoenix) point to a formation history distinct of both the UFDs and CDWs.
Their uniformly ``normal'' carbon abundances, lacking any observed CEMP stars, are consistent with the high gas densities expected in globular cluster environments, which promote efficient metal mixing and comparatively chemically homogeneous star formation.

In summary, a consistent picture is emerging where environments associated with small(er) early star-forming systems have the capability to facilitate and preserve faint supernovae signatures, whereas larger systems are formed from gas dominated by more energetic supernovae.

%%%%%%%%%%%%%%%%%%%%%%%%%%%%%%%%%%%%%%%%%%%%%%%%%%%%
\section{Testing early star formation channels with \ch\ and [Si/H] abundances} \label{sec:dtrans}

\begin{figure*}[p]
    \centering
    \includegraphics[width=6.7in]{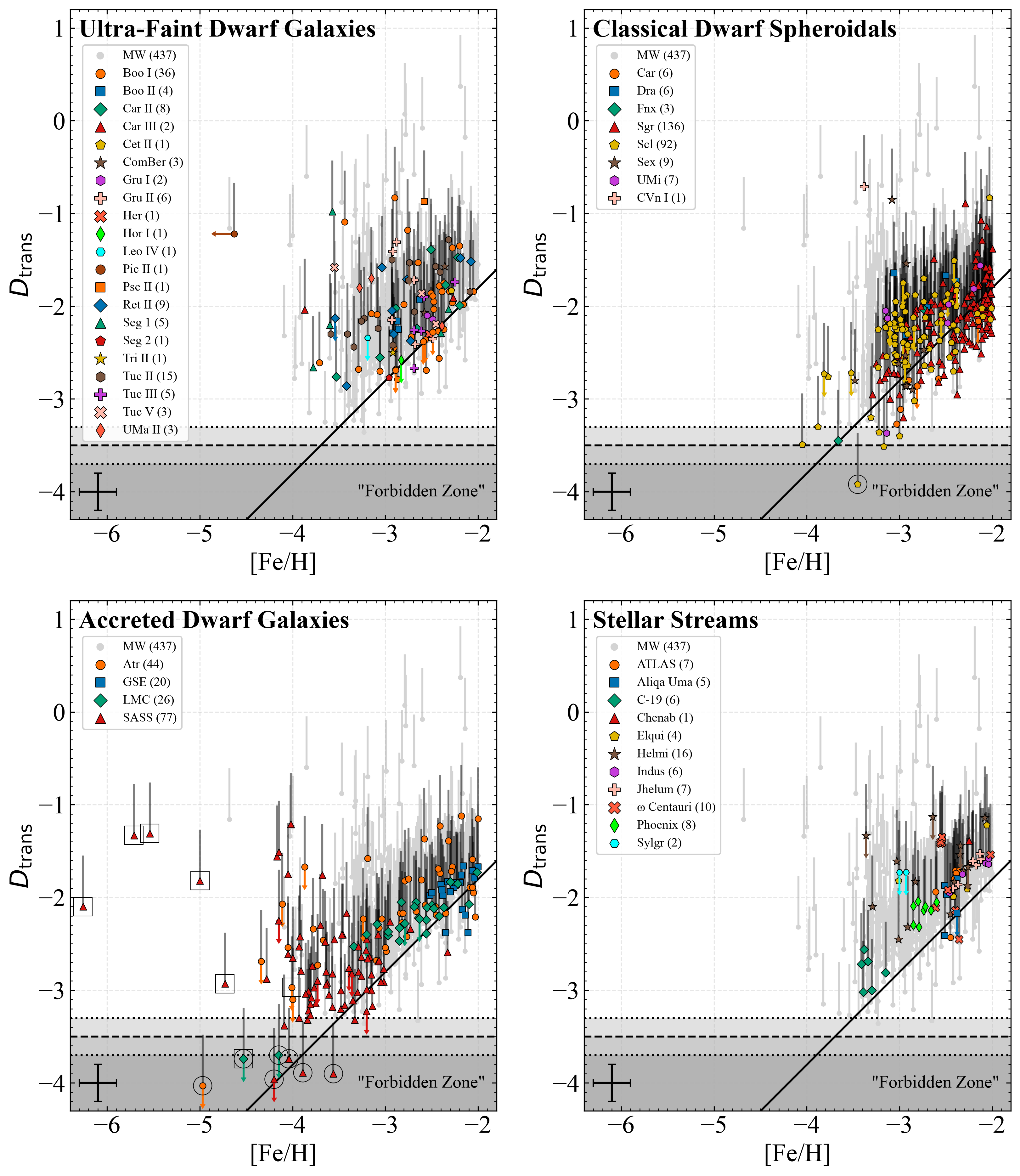}
    \caption{\small
        \feh\ vs $D_{\rm trans}$ for stars in each category of systems, with each panel showing one category, respectively, and the Milky Way stars shown with light gray points for comparison.
        The legends provide information on the individual systems in each category.
        The vertical lines on each data point represent the possible range of the star's $D_{\rm trans}$ value, assuming $-0.7 \le {\rm [C/O]} \le 0.2$.
        The solid diagonal black line represents lower bound of this range using the solar carbon abundances ($\rm{[C/H]} = 0$), scaled by factor of $\rm{[Fe/H]}$ (e.g. $D_{\rm trans, \odot} = {\rm [Fe/H]} \times D_{\rm trans}$).
        The dashed black ($D_{\rm trans} = -3.5$) and dotted black lines ($\pm 0.2$\,dex) represent the limit and associated uncertainty of the ``forbidden zone'' (shaded in gray), where fine-structure line cooling is not efficient for cooling the primordial gas.
        Stars with circles around them have been identified to be in or near the ``forbidden zone'', while stars with squares around them fall in or near dust cooling thresholds (see Figure~\ref{fig:sih_dtrans}).
        Typical error bars of $\pm 0.2$\,dex for each data point are shown in the bottom left of each panel.
    }
    \label{fig:feh_dtrans}
\end{figure*}

The transition from massive, primordial Pop\,III stars to low-mass, second-generation Pop\,II stars required the near-pristine gas to cool and fragment sufficiently to form stars with masses less than $\sim 1 {\rm M}_{\odot}$. 
Metal-line cooling has long been proposed as the dominant cooling channel during this transition, given that Pop\,III supernovae are expected to produce significant CNO yields \citep{Umeda2003a, Heger2010a}.
As a result, \citet{Bromm2003a} predicted that early low-mass Pop\,II stars formed from gas cooled primarily by fine-structure transitions of ${\rm C\,II}$ and ${\rm O\,I}$, rather than molecules that require lower temperatures, or dust that requires much higher densities.

Building on this framework, \citet{Frebel2007c} introduced the transition discriminant, $D_{\rm trans}$, to quantify the combined efficiency of carbon and oxygen fine-structure line cooling and compare it directly with observed metal-poor stellar abundances.\footnote{
    The coefficient of 0.9 was updated by \citealt{Frebel2013a} to reflect revised calculations of the $\rm{O\,I}$ cooling efficiency; the original value introduced by \citet{Frebel2007c} was 0.3.
}
\begin{equation} \label{equ:dtrans}
    D_{\rm trans} \equiv \log_{10}(10^{\rm [C/H]} + 0.9 \times 10^{\rm {[O/H]}}) \gtrsim -3.5
\end{equation}

It was posed that stars formed through this cooling channel should exhibit $D_{\rm trans}$ values above a critical threshold of $D_{\rm trans, crit} = -3.5 \pm 0.2$. 
Conversely, any star with $D_{\rm trans} \lesssim -3.5$ would reside in the so-called ``forbidden zone'', where fine-structure cooling alone is insufficient and the star likely formed through some alternative mechanism, i.e., dust-induced cooling \citep{Schneider2003a, Chiaki2013a, Chiaki2013b}. 
With essentially all of the metal-poor stars known at the time satisfying this criteria, it was concluded that second-generation stars did indeed predominantly form from fine-structure line-cooled gas in the early universe. 

To further test and confirm this picture, we determine the $D_{\rm trans}$ values for all dwarf galaxy and stream stars in our sample. 
Given the general lack of oxygen abundance measurements, we compute a plausible range of $D_{\rm trans}$ values for each star using the corrected \ch\ values and assuming correlated carbon and oxygen production within the range $-0.7 \le {\rm [C/O]} \le 0.2$, following \citep{Spite2005a}.
This range had also previously been used by \citet{Frebel2013a} and \citet{Ji2014a} for an earlier version of this study.
Figure~\ref{fig:feh_dtrans} shows the resulting $D_{\rm trans}$ ranges for each star, represented as vertical black bars.

As expected, the vast majority of stars lie well above the critical threshold, consistent with fine-structure line–cooled gas, or gas that had simply already reached sufficient metallicity for the associated metal cooling.

Interestingly, we identify eight stars (four newly recognized and four previously known) whose $D_{\rm trans}$ ranges lie entirely within or near the ``forbidden zone'', an extremely rare occurrence.
Scl\_002\_06 is a newly recognized ``forbidden'' star in Sculptor \citep{Jablonka2015a}.
Gaia~DR3~4348475068024891776 is a newly discovered ``forbidden'' star with \feh$=-3.6$ associated the SASS population \citep{Hughes2026a}.
SMSS~J184226.25$-$272602.7 is a known ``forbidden'' star (with an upper $D_{\rm trans}$ limit) associated with the SASS population \citep{Jacobson2015a}, for which we have a newly derived carbon measurement (details forthcoming in Yelland et al. 2026b, in preparation).
CS~30336$-$049 is newly recognized ``forbidden'' star associated with the SASS population \citep{Lai2008a}.
LMC-119 is a newly recognized ``forbidden'' star in the LMC \citep{Chiti2024a}.
CD~$-$38$^{\circ}$~245, a well-studied ``forbidden'' star (with an upper $D_{\rm trans}$ limit), is also associated with the SASS population \citep{Cayrel2004a}.
A new spectrum has been acquired to further analysis (details forthcoming in Yelland et al. 2026b, in preparation).
J0715$-$7334, well-known to have the lowest overall metallicity to date, is a recently identified ``forbidden'' star in the LMC \citet{Limberg2025a, Ji2026a}.
Finally, SDSS~J1029+1729, is a known ``forbidden'' star (with an upper $D_{\rm trans}$ limit) associated with the Atari disk \citep{Caffau2011d}.
Additional details are given in Table~\ref{tab:dtrans}.

Given that we can only estimate the oxygen abundance range, we note that for these eight stars to reside above the critical $D_{\rm trans}$ limit, their intrinsic oxygen abundances would need to be at least ${\rm [O/H]} \gtrsim -3.6$, given their respective carbon and iron abundances ($-5 \lesssim$ \feh\ $\lesssim -3.5$).
The oxygen abundances would need to be even higher for the four stars with the carbon upper limits.

In addition to utilizing \ch\ as a cooling mechanism discriminator, the silicon abundances of these stars allows us to perform a similar analysis to identify more concretely whether dust-induced cooling may have played a role.
Silicon acts as a tracer of the influence of dust on the natal gas prior to star formation.
In \citet{Ji2014a}, two dust grain size distributions were tested to determine the critical silicon abundance (${\rm [Si/H]_{crit}}$) required for efficient dust-induced gas fragmentation: a ``standard'' grain size distribution \citep{Pollack1994a, Omukai2010a} and a ``shock'' distribution, in which dust has been processed by a reverse shock following a supernova explosion. 
These models yield two critical silicon thresholds for which dust cooling is no longer viable: ${\rm [Si/H]_{crit,standard}} \le -4.5 \pm 0.1$ and ${\rm [Si/H]_{crit,shock}} \le -5.3 \pm 0.1$.

Silicon abundances are, however, difficult to measure, as one of the two commonly used silicon absorption lines is typically blended with CH lines, and the other line is located in the wing of a hydrogen line.
Hence, not all stars in our sample possess published ${\rm [Si/H]}$ measurements. 
Luckily, eight of the nine ``forbidden zone'' stars and candidates have measured silicon abundances, and only one of them lies near the silicon thresholds at which ``standard'' dust cooling becomes ineffective.
Using the NLTE abundances for this star, J0715$-$7334, found in \citet{Ji2026a}, the star remains within the ``forbidden zone'' but moves out of the forbidden dust cooling region, with $\rm{[Si/H]} \simeq -4.5$.
This suggests that J0715$-$7334, along with the other eight stars, very likely formed from dust-cooled rather than fine-structure-line-cooled early gas.

\input{tables/tab5_dtrans}

\label{tab:dtrans}

\begin{figure*}[!htb]
    \centering
    \includegraphics[width=6.0in]{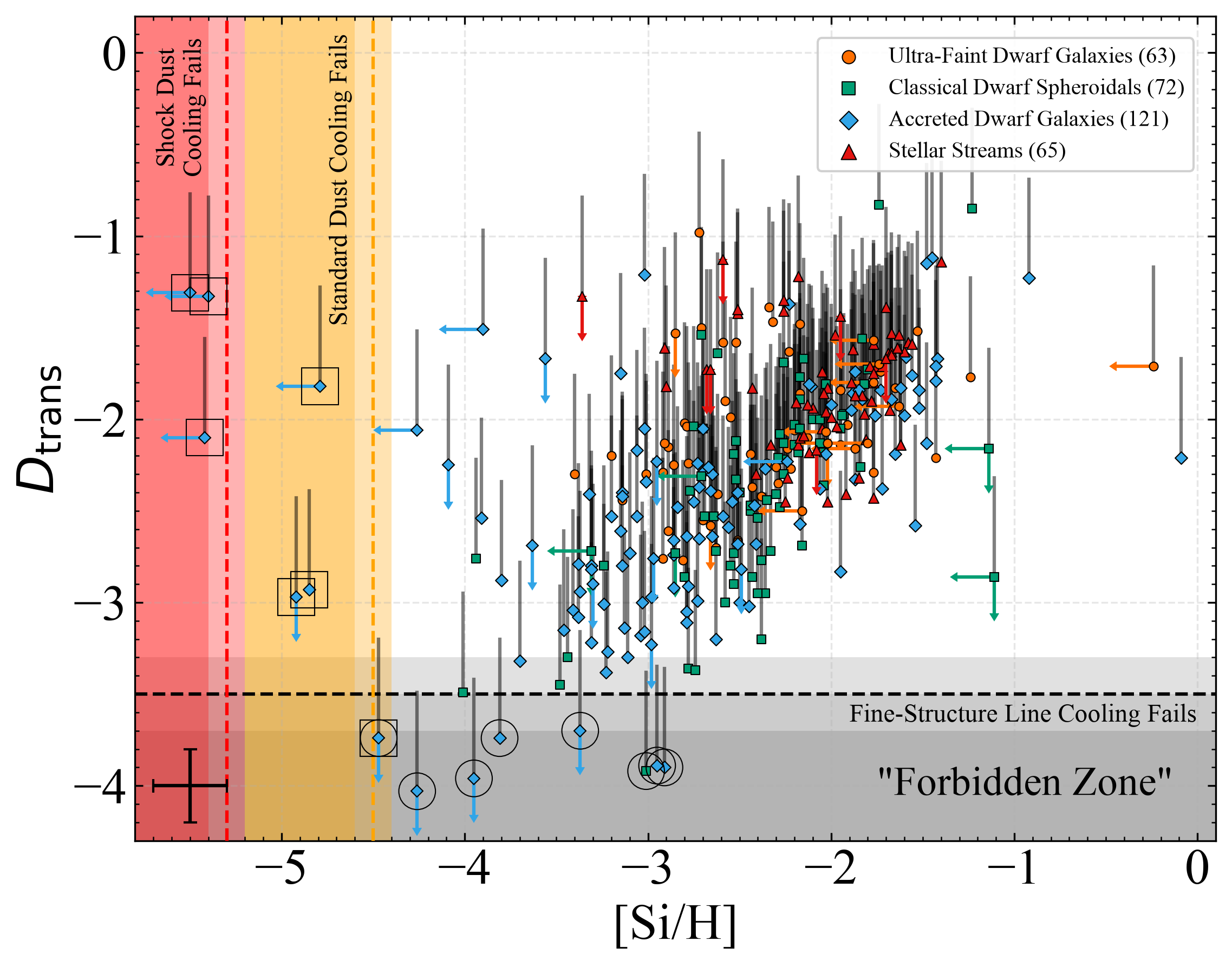}
    \caption{\small
        ${\rm [Si/H]}$ vs $D_{\rm trans}$ for stars in each category of systems.
        The vertical lines on each data point represent the possible range of the star's $D_{\rm trans}$ value, assuming $-0.7 \le {\rm [C/O]} \le 0.2$.
        The vertical orange region (${\rm [Si/H]_{crit,standard}} \le -4.5 \pm 0.1$) and red region (${\rm [Si/H]_{crit,shock}} \le -5.3 \pm 0.1$) represent the thresholds for which ``standard'' and ``shock'' dust cooling remains viable \citep{Ji2014a}.
        Any [Si/H] abundances below these thresholds infer that dust is not efficient for cooling the primordial gas.
        The black dashed line ($D_{\rm trans} = -3.5 \pm 0.2$\,dex) represents the boundary of the ``forbidden zone'' (shaded in gray), where fine-structure line transitions are not also efficient for primordial gas cooling.
        Stars with circles around them have been identified to be in or near the ``forbidden zone'', while stars with squares around them fall in or near dust cooling thresholds.
        Typical error bars of $\pm 0.2$\,dex for each data point are shown in the bottom left of each panel.
    }
    \label{fig:sih_dtrans}
\end{figure*}

Among the remaining stars in the UFD, CDW, ADW, and SS samples with available silicon abundances, we identify six stars with $\rm{[Si/H]}$ values\footnote{For HE~0557$-$4840, HE~1327$-$2326, and HE~0107$-$5240, we adopt the silicon abundances derived in \citealt{Ji2014a}.} below the critical thresholds for dust cooling .
These six objects comprise one Atari star and five SASS stars, as shown in Figure~\ref{fig:sih_dtrans} and listed in Table~\ref{tab:dtrans}.
We note that four of these stars were previously identified in this regime by \citet{Ji2014a}.
Three of these stars fall below the failure threshold for standard dust cooling ($-5.2 \le {\rm [Si/H]} \le -4.6$), while the remaining three lie below the threshold for shock-processed dust (${\rm [Si/H]} \le -5.4$). 
Given their already high $D_{\rm trans}$ values, this confirms that these stars must have formed from fine-structure line cooled gas, rather than dust-cooled gas.

% origins of carbon-poor stars
Having established that dust cooling is not only viable, but perhaps even necessary for the seven ``forbidden zone'' stars and two dust-cooled candidates, we now consider how these objects are distributed across different galactic environments. 

Across both the UFD and SS samples, no ``forbidden zone'' stars or candidates are found; in fact, the lowest values remain entirely above $D_{\rm trans} \sim -3.0$. 
In contrast, two stars in the CDW group reside within or near the ``forbidden zone'', while the other seven dust-cooled stars and candidates are associated with accreted populations, namely the LMC, Atari Disk, and SASS sample.

This demographic pattern, at face value, suggests that early stars forming via dust-induced cooling are predominantly associated with extragalactic or accreted systems.
Given that all other LMC and Sculptor stars exhibit substantially higher $D_{\rm trans}$ values, the two LMC and two Sculptor ``forbidden'' stars and candidates perhaps all formed in a different environment prior to being accreted by their host systems.
By the same analogy, it is plausible that the ``forbidden'' Atari star was also accreted during the early evolution of the Atari system; however, we note that the formation history of the Atari disk remains largely unconstrained, as discussed in Section~\ref{sec:supernovae}.

With the remaining four ``forbidden'' stars and candidates all being in the SASS sample, its not unlikely that the ``forbidden'' stars in the LMC, Scl, and Atari originated in systems more similar to that of the ``forbidden'' SASS stars.
The SASS systems should be very comparable to the UFDs, and in this study, we have focused on the similarities between the high CEMP fractions at the lowest metallicities; however, the lack carbon-poor UFD ``forbidden'' stars infers that at least a subset of SASS stars formed in environments with somewhat different enrichment histories than those of typical UFDs.
Considering that UFDs and SASS stars have some of the highest CEMP fractions, and that the ``forbidden'' stars seem to preferentially originate in these smallest UFD-like systems (inclusive of SASS), we are thus faced with two possible formation pathways:
\begin{enumerate}
    \vspace{-0.2cm}
    \item Early UFD-like environments must be capable of producing both the most carbon-enhanced (relative to iron) and the most carbon-poor (relative to hydrogen) metal-poor stars; or
    \vspace{-0.2cm}
    \item CEMP stars and ``forbidden'' stars must have formed in environments experiencing different enrichment histories.
    \vspace{-0.2cm}
\end{enumerate}

If we consider that UFD stars formed with minimal enrichment from energetic core-collapse events and larger contributions from low-energy faint supernovae (as discussed in Section~\ref{sec:supernovae}), then the ``forbidden'' SASS stars could have formed through an separate, contrasting enrichment channel.
For this rare circumstance to occur, a small UFD-like progenitor would have needed to host at least some high energy core-collapse events and relatively fewer faint supernovae, such that it could remain intact just long enough to form stars before being disrupted or quenched by subsequent enrichment or external factors (Scenario 2 above in Section~\ref{sec:supernovae}).
In this scenario, the stars that formed prior to disruption would be dispersed into the surrounding environments to become part of the present-day halo or be accreted by other satellite systems.

Consequently, one could expect that the ``forbidden'' stars may have more similarities to the stars produced in the smallest of the emerging CDW systems, prior to them accreting any early UFDs.
Hence, we conclude that the second conclusion may be preferred, although simulations and modeling \citep{Brauer2025a} would be needed to further explore this topic. 
More discoveries of these rare ``forbidden'' stars are also needed to more firmly employ them as probes of early galaxy formation channels. 

%%%%%%%%%%%%%%%%%%%%%%%%%%%%%%%%%%%%%%%%%%%%%%%%%%%%
\section{Conclusions and Summary} \label{sec:summary}

Carbon serves as a powerful probe of early chemical enrichment, early star formation and the associated environmental conditions.
To investigate these connections, we compiled a literature sample of carbon abundances for 1045 metal-poor stars spanning 44 systems across four categories of classical dwarf galaxies (CDW), accreted dwarf galaxies (ADW), ultra-faint dwarf galaxies (UFD), and stellar streams (SS), as well as the Milky Way halo (MW). 
Specifically, we established the fraction of CEMP stars with \cfe\ $\ge 0.7$ for each system and category.
Almost all categories show an increase in CEMP fraction with decreasing \feh, except for the CDWs.

% UFD enrichment
The present-day UFDs, with their high CEMP fractions, appear to have been enriched primarily by faint supernovae, which produce high \cfe\ yields, rather than by more energetic core-collapse supernovae or hypernovae, whose \cfe\ yields are closer to the solar ratio \citep{Nomoto2006b, Kobayashi2011e}.
Early UFD-like systems may initially have sampled a broader progenitor explosion-energy distribution; however, only low-mass halos experiencing a truncated distribution, skewed toward lower energies, appear to have survived and continued forming stars.
The broadly similar CEMP fraction behavior among UFDs with sufficient data suggests that this was a common formation channel, likely governed by halo mass and by the inability of these systems to withstand substantial supernova explosion energy injection while sustaining star formation.
How such an explosion-energy distribution maps onto the masses of the first/earliest stars remains to be investigated, but it may offer important clues about the IMF for Pop\,III stars.

\begin{figure*}[!htb]
    \centering
    \includegraphics[width=7in]{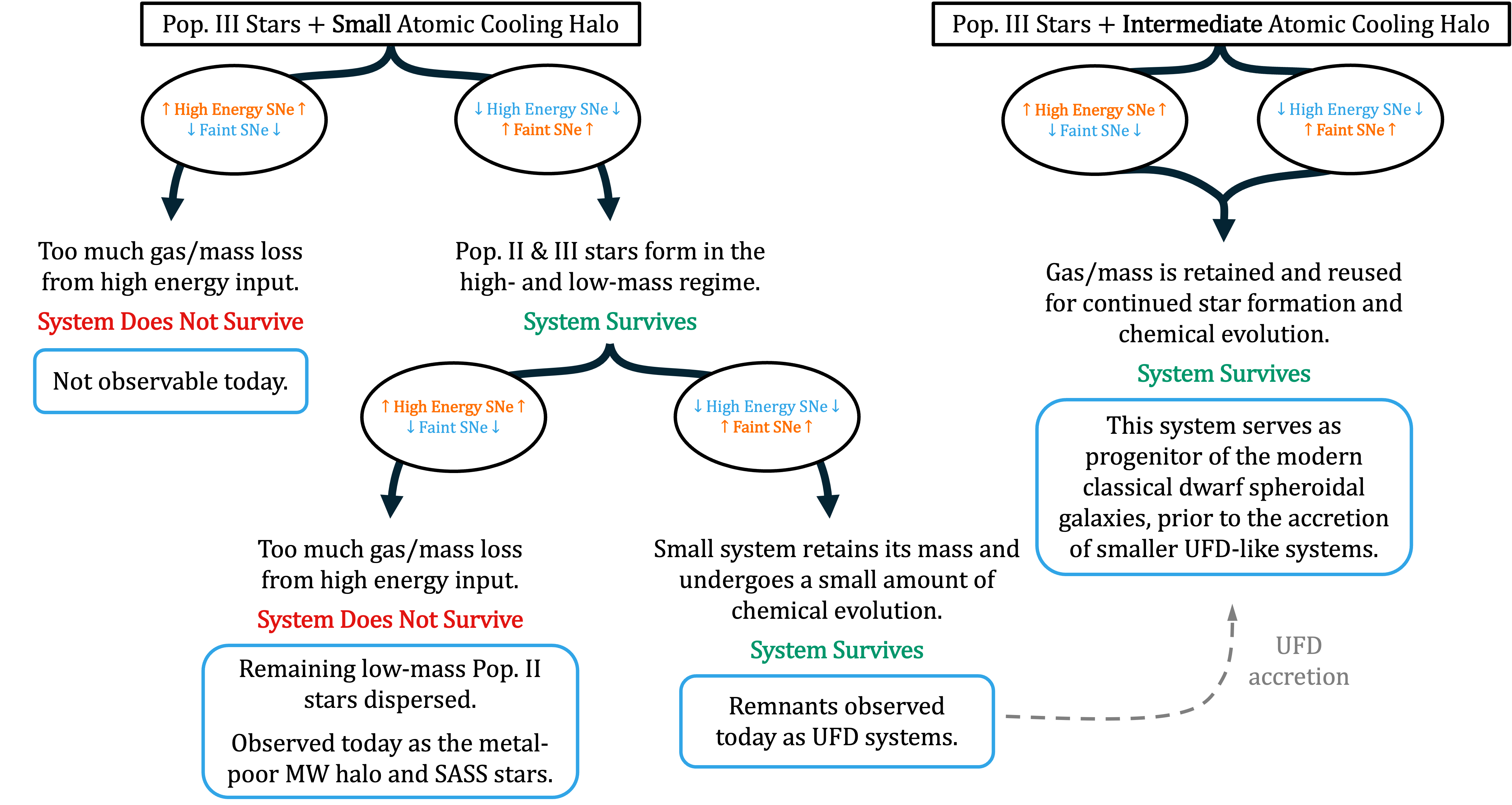}
    \caption{\small
        Flowchart outlining the system survivability as a function of Pop\,III supernova explosion energy. 
        High- and low-energy Pop\,III supernovae influence early dwarf galaxy formation, subsequent star formation, and continued chemical enrichment.
        Depending on the mass of the initial atomic cooling halo, two distinct pathways lead to the formation of UFDs and CDWs, respectively, while a another potentially leads to the formation of some MW halo and SASS stars.
    }
    \label{fig:dwarf_galaxy_formation}
\end{figure*}

% CDW enrichment
In contrast, CDWs may have formed from somewhat larger halos with the capacity to withstand high-energy hypernovae, or more numerous lower-energy supernovae, without quenching star formation, rather than forming primarily through the aggregation of UFD-like systems.
Consequently, these early systems may have hosted massive stars spanning a much broader explosion-energy function, including both faint supernovae and hypernovae, but with a distribution shifted towards the high-energy range, consistent with the low CEMP fractions observed across CDWs.
After their formation, CDWs may have then accreted some of the surviving UFDs at later times, contributing to the small number of CEMP stars observed in them today.
The relationship between the low- and high-energy supernovae and dwarf galaxy survivability is summarized in Figure~\ref{fig:dwarf_galaxy_formation}.

% MCF relationship
Based on this, we uncovered a Magnitude--CEMP fraction (MCF) relation for system containing CEMP stars with \feh\ $\le -2.0$ (see Figure~\ref{fig:mcf_relation}) in which the fainter UFDs have a progressively higher CEMP fraction than the more massive CDWs as a function of absolute $M_{\rm V}$ magnitude.
The MCF relation appears to hold for the entire set of surviving dwarf galaxies investigated in this study (with sufficient data), from the faintest UFDs to the massive LMC, spanning 16 magnitudes and roughly six orders of magnitude in stellar halo mass.
Such a relation predicts the potential discovery of CEMP stars in additional dwarf galaxies as well as their CEMP fractions.
For example, for the SMC and LMC, a $\sim 0\%_{-00}^{+10}$ CEMP fraction is predicted, which thus far is in agreement with the limited data available.
The MCF relation also predicts that UFDs with currently insufficient numbers of observed stars should eventually reveal significant (relative) numbers of CEMP stars, whereas CDWs are expected to host comparatively few.

More generally, this new low-metallicity MCF relation underpins how stellar abundances are powerful indicators of the underlying environmental conditions and properties, including the distributions of progenitor explosion energy, the associated stellar mass of Pop\,III stars, and the halo masses that enabled the survival of the present-day galaxies.
Detailed cosmological simulations, together with additional observations, would be valuable for further testing and refining this relation \citep{Brauer2025a}.
It could also prove useful for constraining galaxy chemical evolution models with respect to the integrated carbon yields of low-metallicity stellar populations.

% [C/H] analysis and dust-cooling candiates
We also investigated the \ch\ abundances of all stars across the four categories to determine their transition discriminants, $D_{\rm trans}$, and thereby probe the conditions of their formation.
Most stars have $D_{\rm trans}$ values well above the critical threshold, suggesting that carbon and oxygen fine-structure line cooling was efficient during early star formation.
But interestingly, we also identified nine stars within or near the so-called ``forbidden zone'' \citep{Frebel2007c, Frebel2013a} where they likely formed via dust cooling in their natal clouds, rather than through fine-structure line cooling.
Seven of these ``forbidden'' stars and candidates are associated with accreted systems (four SASS stars, two LMC stars, one Atari Disk star), while the remaining two objects are associated with Sculptor.

Given that the carbon abundances (and thus, $D_{\rm trans}$ values) of the ``forbidden'' stars and candidates in Sculptor, LMC, and Atari differ significantly from the other stars in their host systems, it is likely they formed in different environments before being accreted.
We suggest that a subset of SASS environments perhaps experienced more energetic or more numerous supernovae, relative to the surviving UFDs, before disrupting or quenching, allowing for stars to form with very low carbon abundances.
As such, small early star-forming environments may follow at least two evolutionary pathways set by their supernova energy input and associated enrichment histories, enabling them to (separately) produce the most carbon-enhanced (relative to iron) or the most carbon-poor (relative to hydrogen) metal-poor stars.
In any case, the existence of these ``forbidden zone'' stars and candidates indicate that dust-induced cooling operated the early universe; however, it was likely a rare channel for forming low-mass second-generation stars, and may have been restricted to particular halo environments with low survival rates, due to e.g. enrichment by energetic hypernovae. 

% Assembly of the MW halo
Finally, in the context of the hierarchical assembly of the Milky Way, the CEMP fraction of halo stars may be less extreme or unusual than previously thought, especially when compared with the aggregate UFD population.
Our results broadly suggest that the old, metal-poor halo (\feh\ $\le -2.0$) was likely assembled from a mixture of early UFD- and CDW-like systems, together with the ADW (including SASS) and SS systems.
Such a composite origin would naturally yield a combined halo CEMP fraction intermediate between the high CEMP fractions typical of UFDs and the lower values seen in CDWs, which is, however, not observed.
Instead, it rather appears that the different stellar populations within the halo were built up separately.

Many of the CEMP halo stars were likely contributed through the accretion of smaller UFD-like systems, whereas the majority of the non-CEMP halo stars came from the more intermediate-sized halos that formed the CDWs.
Towards the lowest metallicities (\feh\ $\le -3.5$), this separation becomes even more stark.
The increased  MW halo CEMP fraction suggests a still greater relative contribution of stars from early UFD-like systems.
On the other hand, towards metallicities of \feh\ $\ge -3.0$, ADW and SS systems likely meaningfully contributed stars to the early halo population.

% The End
Going forward, additional carbon abundance measurements for various metal-poor stars will be essential for refining the CEMP fractions for all known systems (accreted or surviving), improving constraints on the MCF relation between galaxy stellar mass and early chemical enrichment, and further disentangling the roles of different cooling channels in early star formation and dwarf galaxy assembly.
Together, such efforts will further demonstrate how carbon abundances in metal-poor stars reveal distinct star and galaxy formation pathways in the early universe, and eventually provide information to better understand the formation of the galactic halo.

%%%%%%%%%%%%%%%%%%%%%%%%%%%%%%%%%%%%%%%%%%%%%%%%%%%%
\section*{acknowledgments}

% Personal acknowledgements
We thank Ian Roederer for discussion on early carbon production. 
A.Y. thanks Cian Roche, Yadira Gaibor, and Michael Reefe for many discussions on the analysis of this dataset, and also thanks Lana Xu, Jacky Li, and Brandon Gillen for assistance with data processing.
% NSF acknowledgement
A.Y., A.F., and X.O. acknowledge support from NSF-AAG grant AST-2307436.
% NSF (X.O.) acknowledgement
X.O. acknowledge support from NSF under Cooperative Agreement 2421782 and the Simons Foundation grant MPS-AI-00010515 awarded to NSF-Simons AI Institute for Cosmic Origins-CosmicAI. 
% PCLB acknowledgement
A.F. acknowledges support from the PCLB Foundation.
% Gaia acknowledgement
This work has made use of data from the European Space Agency (ESA) mission \textit{Gaia} (\url{https://www.cosmos.esa.int/gaia}), processed by the \textit{Gaia} Data Processing and Analysis Consortium (DPAC, \url{https://www.cosmos.esa.int/web/gaia/dpac/consortium}).
Funding for the DPAC has been provided by national institutions, in particular the institutions participating in the \textit{Gaia} Multilateral Agreement.
% SIMBAD acknowledgement
This research has made use of the SIMBAD database, operated at CDS, Strasbourg, France.
% NASA ADS acknoledgement
Additionally, this research has made use of NASA's Astrophysics Data System Bibliographic Services; the arXiv pre-print server operated by Cornell University; the SIMBAD and VizieR databases hosted by the Strasbourg Astronomical Data Center.

%%%%%%%%%%%%%%%%%%%%%%%%%%%%%%%%%%%%%%%%%%%%%%%%%%%%
\section*{author contribution}

A.Y. was responsible for the data collection, formal analysis, writing the software, administering the project github repositories, along with writing and submitting the manuscript.
A.F. came up with the initial research concept and edited the manuscript.
X.O. helped gather data for the Sagittarius sample.
S.H. assisted in the refinement of the SASS sample.

%%%%%%%%%%%%%%%%%%%%%%%%%%%%%%%%%%%%%%%%%%%%%%%%%%%%
%% Software.
\medskip
\textit{Software:}
\texttt{matplotlib} \citep{Matplotlib2007},
\texttt{numpy} \citep{NumPy2011}
\texttt{scipy} \citep{SciPy2020}
\texttt{astropy} \citep{AstroPy2013, AstroPy2018, AstroPy2022},  
% \texttt{SMHR} \citep{SMHR2014, SMHR2025},
\texttt{pandas} \citep{Pandas2010}

%%%%%%%%%%%%%%%%%%%%%%%%%%%%%%%%%%%%%%%%%%%%%%%%%%%%
%% Bibliography.
% \bibliographystyle{apsrev4-1}
% \bibliography{oja_template}
\bibliography{references, software}{}
\bibliographystyle{aasjournal}

%%%%%%%%%%%%%%%%%%%%%%%%%%%%%%%%%%%%%%%%%%%%%%%%%%%%
%% Appendix.
\begin{appendix}
\input{appendix}
\end{appendix}
\medskip

%%%%%%%%%%%%%%%%%%%%%%%%%%%%%%%%%%%%%%%%%%%%%%%%%%%%
% End of file `main.tex'.
\end{document}

%% file: tables/tab1_sysproperties.tex
\begin{deluxetable*}{llrrr|rrr|rrr}
% \label{tab:sysproperties}
% \tabletypesize{\scriptsize}
\tablecaption{\small
	System properties and their associated stellar populations.This includes the type of system (MW: Milky Way, UFD: Ultra-Faint Dwarf, ADW: Accreted Dwarf, CDW: Classical Dwarf, SS: Stellar Stream), absolute V-band magnitude ($M_{\rm V}$), calculated stellar mass ($M_{\rm stellar}$) in units of $M_{\odot}$, the total number of stars with \feh\ $\le 2.0$ in our sample. For both the ``Observed'' and ``Monte Carlo (MC)'' fractions (see Section~\ref{sec:cempfractions}), we report the number of CEMP stars and total number of stars used in the corresponding CEMP fractions for the \cfe\ $\ge 0.7$ criteria. The references for abundance data of each system can be found in Table~\ref{tab:abunds1}. For the absolute magnitudes, we adopted those collected in \citet{Simon2019a}, \citet{Munoz2018a}, and \citet{Shipp2018a}. The stellar masses of the SS are taken from \citet{Shipp2018a} and \citet{Yuan2022a}, whereas the others are estimated by adopting a constant stellar mass-to-light ratio of one for CDWs and the LMC, and two for UFDs \citep{Simon2019a, Martin2008a}.
}
\tablehead{
	\multicolumn{5}{c}{} & 
	\multicolumn{3}{c}{Observed Fraction} & 
	\multicolumn{3}{c}{MC Fraction} 
	\\
	\colhead{System} & 
	\colhead{Type} & 
	\colhead{$M_{V}$} & 
	\colhead{$M_{\rm stellar}$} & 
	\colhead{$N_{\rm stars}$} & 
	\colhead{$N_{\rm CEMP}^{\rm obs}$} & 
	\colhead{$N_{\rm total}^{\rm obs}$} & 
	\colhead{$f_{\rm CEMP}^{\rm obs} (\%)$} & 
	\colhead{$N_{\rm CEMP}^{\rm mc}$} & 
	\colhead{$N_{\rm total}^{\rm mc}$} & 
	\colhead{$f_{\rm CEMP}^{\rm mc} (\%)$}
}
\startdata
Milky Way (MW)                & MW  & \nodata                    & \nodata                             & 437 & 79 & 437 & 18      & 92 $\pm$ 5 & 437 $\pm$ 0 & ${21}_{-03}^{+03}$ \\
Bootes I (Boo I)              & UFD & ${-6.02}_{-0.25}^{+0.25}$  & ${4.4}_{-0.9}^{+1.1} \times 10^{4}$ & 36  & 8  & 36  & 22      & 8 $\pm$ 1  & 36 $\pm$ 0  & ${22}_{-10}^{+11}$ \\
Bootes II (Boo II)            & UFD & ${-2.94}_{-0.75}^{+0.74}$  & ${2.6}_{-1.3}^{+2.6} \times 10^{3}$ & 4   & 0  & 4   & 0       & 0 $\pm$ 1  & 4 $\pm$ 0   & ${0}_{-00}^{+36}$ \\
Carina II (Car II)            & UFD & ${-4.50}_{-0.10}^{+0.10}$  & ${1.1}_{-0.1}^{+0.1} \times 10^{4}$ & 8   & 1  & 8   & 12      & 2 $\pm$ 1  & 8 $\pm$ 0   & ${25}_{-23}^{+29}$ \\
Carina III (Car III)          & UFD & ${-2.40}_{-0.20}^{+0.20}$  & ${1.6}_{-0.3}^{+0.3} \times 10^{3}$ & 2   & 1  & 2   & 50      & 1 $\pm$ 0  & 2 $\pm$ 0   & ${50}_{-32}^{+32}$ \\
Cetus II (Cet II)             & UFD & ${0.00}_{-0.68}^{+0.68}$   & ${1.7}_{-0.8}^{+1.5} \times 10^{2}$ & 1   & 0  & 1   & 0       & 0 $\pm$ 0  & 1 $\pm$ 0   & ${0}_{-00}^{+62}$ \\
Coma Berenices (ComBer)       & UFD & ${-4.28}_{-0.25}^{+0.25}$  & ${8.8}_{-1.8}^{+2.3} \times 10^{3}$ & 3   & 0  & 3   & 0       & 1 $\pm$ 1  & 3 $\pm$ 0   & ${33}_{-33}^{+50}$ \\
Grus I (Gru I)                & UFD & ${-3.47}_{-0.59}^{+0.59}$  & ${4.2}_{-1.8}^{+3.0} \times 10^{3}$ & 2   & 0  & 2   & 0       & 0 $\pm$ 0  & 2 $\pm$ 0   & ${0}_{-00}^{+39}$ \\
Grus II (Gru II)              & UFD & ${-3.90}_{-0.22}^{+0.22}$  & ${6.2}_{-1.1}^{+1.4} \times 10^{3}$ & 6   & 3  & 6   & 50      & 3 $\pm$ 1  & 6 $\pm$ 0   & ${50}_{-30}^{+30}$ \\
Hercules (Her)                & UFD & ${-5.83}_{-0.17}^{+0.17}$  & ${3.7}_{-0.5}^{+0.6} \times 10^{4}$ & 1   & 0  & 1   & 0       & 0 $\pm$ 0  & 1 $\pm$ 0   & ${0}_{-00}^{+52}$ \\
Horologium I (Hor I)          & UFD & ${-3.76}_{-0.56}^{+0.56}$  & ${5.5}_{-2.2}^{+3.7} \times 10^{3}$ & 1   & 0  & 1   & 0       & 0 $\pm$ 0  & 1 $\pm$ 0   & ${0}_{-00}^{+50}$ \\
Leo IV (Leo IV)               & UFD & ${-4.99}_{-0.26}^{+0.26}$  & ${1.7}_{-0.4}^{+0.5} \times 10^{4}$ & 1   & 0  & 1   & 0       & 0 $\pm$ 0  & 1 $\pm$ 0   & ${0}_{-00}^{+50}$ \\
Pictor II (Pic II)            & UFD & ${-3.20}_{-0.50}^{+0.40}$  & ${3.3}_{-1.0}^{+1.9} \times 10^{3}$ & 1   & 1  & 1   & 100     & 1 $\pm$ 0  & 1 $\pm$ 0   & ${100}_{-50}^{+00}$ \\
Pisces II (Psc II)            & UFD & ${-4.23}_{-0.38}^{+0.38}$  & ${8.4}_{-2.5}^{+3.5} \times 10^{3}$ & 1   & 1  & 1   & 100     & 1 $\pm$ 0  & 1 $\pm$ 0   & ${100}_{-51}^{+00}$ \\
Reticulum II (Ret II)         & UFD & ${-3.88}_{-0.38}^{+0.38}$  & ${6.1}_{-1.8}^{+2.6} \times 10^{3}$ & 9   & 2  & 8   & 25      & 3 $\pm$ 1  & 8 $\pm$ 0   & ${38}_{-25}^{+28}$ \\
Segue 1 (Seg 1)               & UFD & ${-1.30}_{-0.73}^{+0.73}$  & ${5.7}_{-2.8}^{+5.4} \times 10^{2}$ & 5   & 3  & 5   & 60      & 3 $\pm$ 0  & 5 $\pm$ 0   & ${60}_{-29}^{+25}$ \\
Segue 2 (Seg 2)               & UFD & ${-1.98}_{-0.88}^{+0.88}$  & ${1.1}_{-0.6}^{+1.3} \times 10^{3}$ & 1   & 0  & 1   & 0       & 0 $\pm$ 0  & 1 $\pm$ 0   & ${0}_{-00}^{+51}$ \\
Triangulum II (Tri II)        & UFD & ${-1.60}_{-0.76}^{+0.76}$  & ${7.5}_{-3.8}^{+7.6} \times 10^{2}$ & 1   & 0  & 1   & 0       & 0 $\pm$ 0  & 1 $\pm$ 0   & ${0}_{-00}^{+59}$ \\
Tucana II (Tuc II)            & UFD & ${-3.90}_{-0.20}^{+0.20}$  & ${6.2}_{-1.0}^{+1.3} \times 10^{3}$ & 15  & 8  & 14  & 57      & 8 $\pm$ 1  & 14 $\pm$ 0  & ${53}_{-22}^{+22}$ \\
Tucana III (Tuc III)          & UFD & ${-1.49}_{-0.20}^{+0.20}$  & ${6.8}_{-1.1}^{+1.4} \times 10^{2}$ & 5   & 0  & 5   & 0       & 0 $\pm$ 0  & 5 $\pm$ 0   & ${0}_{-00}^{+25}$ \\
Tucana V (Tuc V)              & UFD & ${-1.60}_{-0.49}^{+0.49}$  & ${7.5}_{-2.7}^{+4.3} \times 10^{2}$ & 3   & 1  & 3   & 33      & 1 $\pm$ 0  & 3 $\pm$ 0   & ${33}_{-33}^{+42}$ \\
Ursa Major II (UMa II)        & UFD & ${-4.43}_{-0.26}^{+0.26}$  & ${1.0}_{-0.2}^{+0.3} \times 10^{4}$ & 3   & 2  & 3   & 66      & 2 $\pm$ 0  & 3 $\pm$ 0   & ${67}_{-30}^{+22}$ \\
Canes Venatici I (CVn I)      & CDW & ${-8.73}_{-0.06}^{+0.06}$  & ${2.7}_{-0.1}^{+0.2} \times 10^{5}$ & 1   & 1  & 1   & 100     & 1 $\pm$ 0  & 1 $\pm$ 0   & ${100}_{-50}^{+00}$ \\
Carina (Car)                  & CDW & ${-9.45}_{-0.05}^{+0.05}$  & ${5.2}_{-0.2}^{+0.2} \times 10^{5}$ & 6   & 0  & 6   & 0       & 0 $\pm$ 0  & 6 $\pm$ 0   & ${0}_{-00}^{+14}$ \\
Draco (Dra)                   & CDW & ${-8.88}_{-0.05}^{+0.05}$  & ${3.1}_{-0.1}^{+0.1} \times 10^{5}$ & 6   & 1  & 6   & 16      & 2 $\pm$ 1  & 6 $\pm$ 0   & ${33}_{-28}^{+33}$ \\
Fornax (Fnx)                  & CDW & ${-13.34}_{-0.14}^{+0.14}$ & ${1.9}_{-0.2}^{+0.3} \times 10^{7}$ & 3   & 0  & 3   & 0       & 0 $\pm$ 0  & 3 $\pm$ 0   & ${0}_{-00}^{+38}$ \\
Sagittarius (Sgr)             & CDW & ${-13.50}_{-0.15}^{+0.15}$ & ${2.2}_{-0.3}^{+0.3} \times 10^{7}$ & 136 & 5  & 136 & 3       & 7 $\pm$ 2  & 136 $\pm$ 0 & ${5}_{-03}^{+04}$ \\
Sculptor (Scl)                & CDW & ${-10.82}_{-0.14}^{+0.14}$ & ${1.8}_{-0.2}^{+0.3} \times 10^{6}$ & 92  & 14 & 91  & 15      & 15 $\pm$ 3 & 90 $\pm$ 1  & ${16}_{-06}^{+07}$ \\
Sextans (Sex)                 & CDW & ${-8.94}_{-0.06}^{+0.06}$  & ${3.2}_{-0.2}^{+0.2} \times 10^{5}$ & 9   & 2  & 9   & 22      & 2 $\pm$ 0  & 9 $\pm$ 0   & ${22}_{-15}^{+20}$ \\
Ursa Minor (UMi)              & CDW & ${-9.03}_{-0.05}^{+0.05}$  & ${3.5}_{-0.2}^{+0.2} \times 10^{5}$ & 7   & 2  & 7   & 28      & 2 $\pm$ 1  & 7 $\pm$ 0   & ${29}_{-23}^{+28}$ \\
Atari Disk (Atr)              & ADW & \nodata                    & \nodata                             & 44  & 11 & 41  & 26      & 11 $\pm$ 2 & 40 $\pm$ 1  & ${28}_{-11}^{+12}$ \\
Gaia-Sausage/Enceladus (GSE)  & ADW & \nodata                    & \nodata                             & 20  & 0  & 20  & 0       & 0 $\pm$ 0  & 20 $\pm$ 0  & ${0}_{-00}^{+07}$ \\
Large Magellanic Cloud (LMC)  & ADW & ${-18.10}_{-0.10}^{+0.10}$ & ${1.5}_{-0.1}^{+0.1} \times 10^{9}$ & 26  & 0  & 26  & 0       & 1 $\pm$ 1  & 26 $\pm$ 0  & ${4}_{-03}^{+10}$ \\
SASS                          & ADW & \nodata                    & \nodata                             & 77  & 24 & 76  & 31      & 27 $\pm$ 2 & 76 $\pm$ 0  & ${36}_{-08}^{+08}$ \\
ATLAS                         & SS  & ${-4.50}_{-0.40}^{+0.40}$  & ${7.4} \times 10^{3}$               & 7   & 0  & 7   & 0       & 0 $\pm$ 1  & 7 $\pm$ 0   & ${0}_{-00}^{+23}$ \\
Aliqa Uma                     & SS  & ${-3.40}_{-0.40}^{+0.40}$  & ${2.3} \times 10^{3}$               & 5   & 0  & 5   & 0       & 0 $\pm$ 0  & 5 $\pm$ 0   & ${0}_{-00}^{+25}$ \\
C-19                          & SS  & \nodata                    & ${1.0} \times 10^{4}$               & 6   & 0  & 6   & 0       & 1 $\pm$ 1  & 6 $\pm$ 0   & ${17}_{-16}^{+31}$ \\
Chenab                        & SS  & ${-5.70}_{-0.40}^{+0.40}$  & ${1.8} \times 10^{4}$               & 1   & 0  & 1   & 0       & 0 $\pm$ 0  & 1 $\pm$ 0   & ${0}_{-00}^{+99}$ \\
Elqui                         & SS  & ${-4.90}_{-0.40}^{+0.40}$  & ${1.0} \times 10^{4}$               & 4   & 1  & 4   & 25      & 1 $\pm$ 1  & 4 $\pm$ 0   & ${25}_{-25}^{+39}$ \\
Helmi                         & SS  & \nodata                    & \nodata                             & 16  & 4  & 14  & 28      & 4 $\pm$ 1  & 14 $\pm$ 0  & ${29}_{-18}^{+20}$ \\
Indus                         & SS  & ${-6.20}_{-0.40}^{+0.40}$  & ${3.4} \times 10^{4}$               & 6   & 0  & 6   & 0       & 0 $\pm$ 0  & 6 $\pm$ 0   & ${0}_{-00}^{+21}$ \\
Jhelum                        & SS  & ${-5.10}_{-0.40}^{+0.40}$  & ${1.3} \times 10^{4}$               & 7   & 0  & 7   & 0       & 0 $\pm$ 1  & 7 $\pm$ 0   & ${0}_{-00}^{+20}$ \\
Omega-Centauri ($\omega$ Cen) & SS  & \nodata                    & \nodata                             & 10  & 4  & 10  & 40      & 4 $\pm$ 1  & 10 $\pm$ 0  & ${40}_{-21}^{+22}$ \\
Phoenix                       & SS  & ${-3.60}_{-0.40}^{+0.40}$  & ${2.8} \times 10^{3}$               & 8   & 0  & 8   & 0       & 1 $\pm$ 1  & 8 $\pm$ 0   & ${13}_{-12}^{+26}$ \\
Sylgr                         & SS  & \nodata                    & \nodata                             & 2   & 0  & 0   & \nodata & 0 $\pm$ 0  & 0 $\pm$ 1   & ${0}_{-00}^{+50}$
\enddata
\end{deluxetable*}
% -------------------------------- 
\nocite{Munoz2018a} 
\nocite{Shipp2018a} 
\nocite{Simon2019a} 
\nocite{Yuan2022a} 
% -------------------------------- 
\nocite{Aguado2017a} 
\nocite{Aguado2022a} 
\nocite{Akerman2004a} 
\nocite{Aoki2005c} 
\nocite{Aoki2006b} 
\nocite{Aoki2007b} 
\nocite{Aoki2008a} 
\nocite{Aoki2013a} 
\nocite{Barklem2005b} 
\nocite{Bonifacio2009b} 
\nocite{Caffau2011d} 
\nocite{Casey2017b} 
\nocite{Cayrel2004a} 
\nocite{Chiti2018a} 
\nocite{Chiti2018b} 
\nocite{Chiti2023a} 
\nocite{Chiti2024a} 
\nocite{Chiti2025a} 
\nocite{Cohen2007a} 
\nocite{Cohen2008a} 
\nocite{Cohen2009a} 
\nocite{Cohen2010a} 
\nocite{Cohen2013a} 
\nocite{Ezzeddine2020a} 
\nocite{Francois2007a} 
\nocite{Francois2016a} 
\nocite{Francois2018b} 
\nocite{Frebel2007a} 
\nocite{Frebel2008a} 
\nocite{Frebel2010a} 
\nocite{Frebel2010b} 
\nocite{Frebel2013c} 
\nocite{Frebel2014a} 
\nocite{Frebel2015a} 
\nocite{Frebel2016a} 
\nocite{Fulbright2004a} 
\nocite{Gilmore2013a} 
\nocite{Gull2021a} 
\nocite{HansenC2018a} 
\nocite{HansenT2014a} 
\nocite{HansenT2015a} 
\nocite{HansenT2017a} 
\nocite{HansenT2018a} 
\nocite{HansenT2020a} 
\nocite{HansenT2024a} 
\nocite{Hollek2011a} 
\nocite{Holmbeck2020a} 
\nocite{Honda2004a} 
\nocite{Hughes2026a} 
\nocite{Ishigaki2014a} 
\nocite{Jablonka2015a} 
\nocite{Jacobson2015a} 
\nocite{Ji2016a} 
\nocite{Ji2016c} 
\nocite{Ji2018a} 
\nocite{Ji2019a} 
\nocite{Ji2020a} 
\nocite{Ji2020b} 
\nocite{Ji2026a} 
\nocite{Keller2014a} 
\nocite{Kirby2012c} 
\nocite{Kirby2017b} 
\nocite{Lai2007a} 
\nocite{Lai2008a} 
\nocite{Lai2011b} 
\nocite{Li2015a} 
\nocite{Lucchesi2020a} 
\nocite{Lucchesi2024a} 
\nocite{Lucey2026b} 
\nocite{Mardini2022b} 
\nocite{Mardini2024b} 
\nocite{Marshall2019a} 
\nocite{Martin2022a} 
\nocite{Masseron2010a} 
\nocite{Masseron2012a} 
\nocite{McWilliam1995a} 
\nocite{Melendez2002a} 
\nocite{Nagasawa2018a} 
\nocite{Nordlander2019a} 
\nocite{Norris2007a} 
\nocite{Norris2010a} 
\nocite{Norris2010b} 
\nocite{Norris2010c} 
\nocite{Ou2024c} 
\nocite{Ou2025a} 
\nocite{Placco2014a} 
\nocite{Placco2015a} 
\nocite{Placco2016b} 
\nocite{Preston2006a} 
\nocite{Roederer2010a} 
\nocite{Roederer2010a} 
\nocite{Roederer2014a} 
\nocite{Roederer2014b} 
\nocite{Roederer2014c} 
\nocite{Roederer2016b} 
\nocite{Roederer2019d} 
\nocite{Roederer2023a} 
\nocite{Ruchti2011a} 
\nocite{Sakari2018b} 
\nocite{Sbordone2020a} 
\nocite{Sestito2024b} 
\nocite{Sestito2024d} 
\nocite{Simmerer2004a} 
\nocite{Simon2010a} 
\nocite{Simon2015a} 
\nocite{Skuladottir2015a} 
\nocite{Skuladottir2024a} 
\nocite{Sneden2003d} 
\nocite{Spite2013a} 
\nocite{Spite2018a} 
\nocite{Starkenburg2013a} 
\nocite{Tafelmeyer2010a} 
\nocite{Venn2012a} 
\nocite{Verdiani2026a} 
\nocite{Webber2023a} 
\nocite{Yong2013a} 
\nocite{Yoon2020a} 
\nocite{Zhang2011a} 
% --------------------------------

%% file: tables/tab2_abundances1.tex
% Table 2: Abundance Table (excerpt), Part 1
\begin{deluxetable*}{llllllrrr}[htb!]
% \label{tab:abunds_stub_1}
% \tabletypesize{\scriptsize}
\tablecaption{\small
    Stellar abundances of stars in the dwarf galaxies, streams, and Milky Way halo. Solar abundances adopted are from \citet{Asplund2009a}. Due to the number of columns, the table has been broken up below. Furthermore, this table is available in its entirety in machine-readable forms in the online journal. A portion is shown here for guidance regarding its form and content.
}
\tablehead{
	\colhead{Star name} & 
	\colhead{Query ID} & 
	\colhead{System} & 
	\colhead{Class} & 
	\colhead{R.A.} & 
	\colhead{Decl.} & 
	\colhead{$\log g$} & 
	\colhead{$A({\rm Fe})$} & 
	\colhead{$\rm{[Fe/H]}$}
}
\startdata
ATLAS\_1                     & Gaia DR3 2349268564550587904  & ATLAS                  & \nodata    & 00:48:55.0 & $-$22:44:58.0 & 1.97 & 5.10    & $-$2.40 \\
ATLAS\_27                    & Gaia DR3 2346224467824940544  & ATLAS                  & rI         & 00:52:59.3 & $-$22:54:15.5 & 1.86 & 5.14    & $-$2.36 \\
ATLAS\_0                     & Gaia DR3 2345957664457105408  & ATLAS                  & rI         & 00:58:40.1 & $-$23:51:49.7 & 1.52 & 5.09    & $-$2.41 \\
ATLAS\_26                    & Gaia DR3 5040976937490509184  & ATLAS                  & rI         & 01:11:13.0 & $-$24:44:48.3 & 1.75 & 5.05    & $-$2.45 \\
ATLAS\_25                    & Gaia DR3 5040671754294144512  & ATLAS                  & rI         & 01:12:21.8 & $-$25:44:52.2 & 1.75 & 5.06    & $-$2.44 \\
ATLAS\_22                    & Gaia DR3 5039838702437479936  & ATLAS                  & rI         & 01:16:27.1 & $-$26:07:01.0 & 1.51 & 5.11    & $-$2.39 \\
ATLAS\_12                    & Gaia DR3 5022844307121290752  & ATLAS                  & rI         & 01:40:08.7 & $-$29:52:14.6 & 1.16 & 4.89    & $-$2.61 \\
AliqaUma\_9                  & Gaia DR3 4971176778264340352  & Aliqa Uma              & rI         & 02:09:08.3 & $-$32:46:06.1 & 1.14 & 4.99    & $-$2.51 \\
AliqaUma\_10                 & Gaia DR3 4971328167270778496  & Aliqa Uma              & rI         & 02:09:58.9 & $-$32:05:40.0 & 1.45 & 5.12    & $-$2.38 \\
AliqaUma\_7                  & Gaia DR3 4969961611757057536  & Aliqa Uma              & rI         & 02:16:18.9 & $-$34:06:22.9 & 1.82 & 4.99    & $-$2.51 \\
AliqaUma\_5                  & Gaia DR3 4966915105554905344  & Aliqa Uma              & rI         & 02:26:26.2 & $-$35:22:26.1 & 1.13 & 5.01    & $-$2.49 \\
AliqaUma\_0                  & Gaia DR3 4953695608534281088  & Aliqa Uma              & \nodata    & 02:35:26.1 & $-$37:22:30.2 & 1.97 & 5.12    & $-$2.38 \\
HE~0023$-$4825               & HE 0023-4825                  & Atari                  & \nodata    & 00:25:50.3 & $-$48:08:27.0 & 3.63 & 5.44    & $-$2.06 \\
HD~2796                      & HD 2796                       & Atari                  & \nodata    & 00:31:16.9 & $-$16:47:40.8 & 1.50 & 5.05    & $-$2.45 \\
CD$-$54$^{\circ}$~503        & CD-54 503                     & Atari                  & rII        & 02:21:55.6 & $-$54:10:14.5 & 3.53 & 5.41    & $-$2.09 \\
\enddata
\end{deluxetable*}

%% file: tables/tab3_abundances2.tex
% Table 2: Abundance Table (excerpt), Part 2
\begin{deluxetable*}{rrrrrrrrrrrl}[htb!]
% \label{tab:abunds_stub_2}
% \tabletypesize{\scriptsize}
\tablecaption{Stellar abundances continued.}
\tablehead{
	\colhead{$A({\rm C})$} & 
	\colhead{$\Delta A({\rm C})$} & 
	\colhead{$A({\rm C})_{\rm corr}$} & 
	\colhead{$\rm{[C/H]}$} & 
	\colhead{$\rm{[C/H]}_{\rm corr}$} & 
	\colhead{$\rm{[C/Fe]}$} & 
	\colhead{$\rm{[C/Fe]}_{\rm corr}$} & 
	\colhead{$\rm{[Si/H]}$} & 
	\colhead{$\rm{[Ba/H]}$} & 
	\colhead{$\rm{[Sr/H]}$} & 
	\colhead{$\rm{[Eu/H]}$} & 
	\colhead{Reference}
}
\startdata
6.43    & $+$0.09 & 6.52    & $-$2.00  & $-$1.91  & 0.40     & 0.49     & $-$1.75  & $-$2.69  & $-$2.67  & $<-$1.64 & \citet{Ji2020b} \\
6.37    & $+$0.15 & 6.52    & $-$2.06  & $-$1.91  & 0.30     & 0.45     & $-$1.72  & $-$2.59  & $-$2.45  & $-$2.01  & \citet{Ji2020b} \\
5.96    & $+$0.47 & 6.43    & $-$2.47  & $-$2.00  & $-$0.06  & 0.41     & $-$1.89  & $-$2.42  & $-$2.23  & $-$2.05  & \citet{Ji2020b} \\
5.58    & $+$0.22 & 5.80    & $-$2.85  & $-$2.63  & $-$0.40  & $-$0.18  & $-$1.77  & $-$2.24  & $-$2.01  & $-$2.00  & \citet{Ji2020b} \\
6.04    & $+$0.23 & 6.27    & $-$2.39  & $-$2.16  & 0.05     & 0.28     & $-$2.03  & $-$2.48  & $-$2.21  & $-$1.91  & \citet{Ji2020b} \\
6.00    & $+$0.49 & 6.49    & $-$2.43  & $-$1.94  & $-$0.04  & 0.45     & $-$2.05  & $-$2.27  & $-$2.14  & $-$2.07  & \citet{Ji2020b} \\
5.62    & $+$0.67 & 6.29    & $-$2.81  & $-$2.14  & $-$0.20  & 0.47     & $-$2.10  & $-$2.66  & $-$2.26  & $-$2.14  & \citet{Ji2020b} \\
5.68    & $+$0.68 & 6.36    & $-$2.75  & $-$2.07  & $-$0.24  & 0.44     & $-$1.83  & $-$2.41  & $-$2.28  & $-$2.18  & \citet{Ji2020b} \\
5.92    & $+$0.51 & 6.43    & $-$2.51  & $-$2.00  & $-$0.13  & 0.38     & $-$1.77  & $-$2.33  & $-$2.16  & $-$1.99  & \citet{Ji2020b} \\
5.62    & $+$0.20 & 5.82    & $-$2.81  & $-$2.61  & $-$0.30  & $-$0.10  & $-$1.92  & $-$2.39  & $-$2.29  & $-$2.07  & \citet{Ji2020b} \\
5.57    & $+$0.69 & 6.26    & $-$2.86  & $-$2.17  & $-$0.37  & 0.32     & $-$1.82  & $-$2.44  & $-$2.20  & $-$2.05  & \citet{Ji2020b} \\
$<$6.02 & $+$0.04 & $<$6.06 & $<-$2.41 & $<-$2.37 & $<-$0.03 & $<$0.01  & $-$2.08  & $-$2.63  & $-$2.87  & $<-$1.65 & \citet{Ji2020b} \\
6.64    & $+$0.00 & 6.64    & $-$1.79  & $-$1.79  & 0.27     & 0.27     & \nodata  & $-$2.17  & $-$1.98  & \nodata  & \citet{Barklem2005b} \\
5.55    & $+$0.49 & 6.04    & $-$2.88  & $-$2.39  & $-$0.43  & 0.06     & $-$2.03  & \nodata  & \nodata  & \nodata  & \citet{Cayrel2004a} \\
6.34    & $+$0.00 & 6.34    & $-$2.09  & $-$2.09  & 0.00     & 0.00     & \nodata  & $-$1.67  & $-$1.87  & $-$1.29  & \citet{Holmbeck2020a} \\
\enddata
\end{deluxetable*}

%% file: tables/tab4_cempfractions.tex
\begin{deluxetable*}{crcrc|rcrc}[p]
% \label{tab:cempfractions}
% \tabletypesize{\scriptsize}
\tablewidth{0.85\textwidth}
\tablecaption{\small
	Cumulative CEMP star fractions of each type of system based on two thresholds: \cfe\ $\ge 0.7$ and \cfe\ $\ge 1.0$. The ``Observed Fraction'' percentages and CEMP star fractions are found from the reported literature \cfe\ abundances, whereas the ``MC Fraction'' values are calculated by accounting for the uncertainties in the CEMP star classification and sample size.
}
\tablehead{
	\colhead{} & 
	\multicolumn{2}{c}{Observed Fraction} & 
	\multicolumn{2}{c}{MC Fraction} & 
	\multicolumn{2}{c}{Observed Fraction} & 
	\multicolumn{2}{c}{MC Fraction}
	\\
	\colhead{$\rm{[Fe/H]} \le$} & 
	\multicolumn{2}{c}{$\rm{[C/Fe]} \ge 0.7$} & 
	\multicolumn{2}{c}{$\rm{[C/Fe]}_{\rm mc} \ge 0.7$} & 
	\multicolumn{2}{c}{$\rm{[C/Fe]} \ge 1.0$} & 
	\multicolumn{2}{c}{$\rm{[C/Fe]}_{\rm mc} \ge 1.0$}
}
\startdata
\multicolumn{9}{c}{Milky Way} \\
\hline
\multicolumn{1}{c|}{$-$2.0} & 18 & (79/437) & $21_{-03}^{+03}$ & (92/437) & 12 & (51/437) & $12_{-02}^{+02}$ & (51/437) \\
\multicolumn{1}{c|}{$-$2.5} & 22 & (73/327) & $26_{-04}^{+04}$ & (84/327) & 15 & (48/327) & $14_{-03}^{+03}$ & (47/327) \\
\multicolumn{1}{c|}{$-$3.0} & 44 & (52/117) & $44_{-07}^{+07}$ & (51/117) & 32 & (37/117) & $30_{-06}^{+06}$ & (35/117) \\
\multicolumn{1}{c|}{$-$3.5} & 80 & (12/15) & $80_{-17}^{+13}$ & (12/15) & 73 & (11/15) & $67_{-20}^{+18}$ & (10/15) \\
\multicolumn{1}{c|}{$-$4.0} & 100 & (4/4) & $100_{-20}^{+00}$ & (4/4) & 100 & (4/4) & $100_{-21}^{+00}$ & (4/4) \\
\multicolumn{1}{c|}{$-$4.5} & 100 & (1/1) & $100_{-50}^{+00}$ & (1/1) & 100 & (1/1) & $100_{-50}^{+00}$ & (1/1) \\
\multicolumn{1}{c|}{$-$5.0} & \nodata & \nodata & \nodata & \nodata & \nodata & \nodata & \nodata & \nodata \\
\hline
\multicolumn{9}{c}{Ultra-Faint Dwarf Galaxies} \\
\hline
\multicolumn{1}{c|}{$-$2.0} & 29 & (31/107) & $31_{-07}^{+07}$ & (34/107) & 15 & (16/108) & $18_{-05}^{+06}$ & (19/108) \\
\multicolumn{1}{c|}{$-$2.5} & 40 & (29/72) & $42_{-09}^{+09}$ & (30/72) & 22 & (16/73) & $26_{-07}^{+08}$ & (19/73) \\
\multicolumn{1}{c|}{$-$3.0} & 72 & (18/25) & $68_{-15}^{+14}$ & (17/25) & 40 & (10/25) & $44_{-15}^{+16}$ & (11/25) \\
\multicolumn{1}{c|}{$-$3.5} & 89 & (8/9) & $89_{-22}^{+11}$ & (8/9) & 67 & (6/9) & $67_{-27}^{+23}$ & (6/9) \\
\multicolumn{1}{c|}{$-$4.0} & 100 & (1/1) & $100_{-50}^{+00}$ & (1/1) & 100 & (1/1) & $100_{-50}^{+00}$ & (1/1) \\
\multicolumn{1}{c|}{$-$4.5} & 100 & (1/1) & $100_{-50}^{+00}$ & (1/1) & 100 & (1/1) & $100_{-50}^{+00}$ & (1/1) \\
\multicolumn{1}{c|}{$-$5.0} & \nodata & \nodata & \nodata & \nodata & \nodata & \nodata & \nodata & \nodata \\
\hline
\multicolumn{9}{c}{Classical Dwarf Spheroidal Galaxies} \\
\hline
\multicolumn{1}{c|}{$-$2.0} & 10 & (25/259) & $11_{-03}^{+03}$ & (29/258) & 2 & (6/260) & $3_{-02}^{+02}$ & (9/260) \\
\multicolumn{1}{c|}{$-$2.5} & 14 & (21/145) & $17_{-05}^{+05}$ & (24/145) & 3 & (4/146) & $5_{-03}^{+03}$ & (7/146) \\
\multicolumn{1}{c|}{$-$3.0} & 24 & (13/55) & $24_{-09}^{+10}$ & (13/54) & 5 & (3/55) & $9_{-06}^{+07}$ & (5/55) \\
\multicolumn{1}{c|}{$-$3.5} & 17 & (1/6) & $17_{-16}^{+30}$ & (1/6) & 0 & (0/6) & $0_{-00}^{+21}$ & (0/6) \\
\multicolumn{1}{c|}{$-$4.0} & 0 & (0/1) & $0_{-00}^{+71}$ & (0/1) & 0 & (0/1) & $0_{-00}^{+53}$ & (0/1) \\
\multicolumn{1}{c|}{$-$4.5} & \nodata & \nodata & \nodata & \nodata & \nodata & \nodata & \nodata & \nodata \\
% \multicolumn{1}{c|}{$-$5.0} & \nodata & \nodata & \nodata & \nodata & \nodata & \nodata & \nodata & \nodata \\
\hline
\multicolumn{9}{c}{Accreted Dwarf Galaxies} \\
\hline
\multicolumn{1}{c|}{$-$2.0} & 21 & (35/163) & $24_{-05}^{+05}$ & (39/162) & 14 & (23/163) & $15_{-04}^{+04}$ & (24/163) \\
\multicolumn{1}{c|}{$-$2.5} & 27 & (32/120) & $30_{-06}^{+07}$ & (36/119) & 19 & (23/120) & $19_{-05}^{+05}$ & (23/120) \\
\multicolumn{1}{c|}{$-$3.0} & 32 & (29/91) & $36_{-07}^{+08}$ & (32/90) & 24 & (22/91) & $24_{-06}^{+07}$ & (22/91) \\
\multicolumn{1}{c|}{$-$3.5} & 48 & (25/52) & $52_{-10}^{+10}$ & (27/51) & 38 & (20/52) & $38_{-09}^{+10}$ & (20/52) \\
\multicolumn{1}{c|}{$-$4.0} & 64 & (14/22) & $67_{-14}^{+12}$ & (14/21) & 64 & (14/22) & $62_{-14}^{+13}$ & (14/22) \\
\multicolumn{1}{c|}{$-$4.5} & 75 & (6/8) & $75_{-23}^{+17}$ & (6/8) & 75 & (6/8) & $75_{-19}^{+13}$ & (6/8) \\
\multicolumn{1}{c|}{$-$5.0} & 100 & (5/5) & $100_{-17}^{+00}$ & (5/5) & 100 & (5/5) & $100_{-17}^{+00}$ & (5/5) \\
\hline
\multicolumn{9}{c}{Accreted Dwarf Galaxies (without SASS)} \\
\hline
\multicolumn{1}{c|}{$-$2.0} & 13 & (11/87) & $15_{-06}^{+07}$ & (13/86) & 6 & (5/87) & $6_{-04}^{+05}$ & (5/87) \\
\multicolumn{1}{c|}{$-$2.5} & 18 & (8/45) & $21_{-09}^{+11}$ & (9/44) & 11 & (5/45) & $9_{-06}^{+08}$ & (4/45) \\
\multicolumn{1}{c|}{$-$3.0} & 28 & (5/18) & $35_{-17}^{+19}$ & (6/17) & 22 & (4/18) & $21_{-12}^{+15}$ & (4/18) \\
\multicolumn{1}{c|}{$-$3.5} & 44 & (4/9) & $44_{-23}^{+24}$ & (4/8) & 33 & (3/9) & $30_{-20}^{+24}$ & (3/9) \\
\multicolumn{1}{c|}{$-$4.0} & 17 & (1/6) & $20_{-16}^{+26}$ & (1/5) & 17 & (1/6) & $17_{-14}^{+24}$ & (1/6) \\
\multicolumn{1}{c|}{$-$4.5} & 0 & (0/2) & $0_{-00}^{+33}$ & (0/2) & 0 & (0/2) & $0_{-00}^{+33}$ & (0/2) \\
\multicolumn{1}{c|}{$-$5.0} & \nodata & \nodata & \nodata & \nodata & \nodata & \nodata & \nodata & \nodata \\
\hline
\multicolumn{9}{c}{Accreted Dwarf Galaxies (only SASS)} \\
\hline
\multicolumn{1}{c|}{$-$2.0} & 32 & (24/76) & $36_{-08}^{+08}$ & (27/76) & 24 & (18/76) & $25_{-07}^{+07}$ & (19/76) \\
\multicolumn{1}{c|}{$-$2.5} & 32 & (24/75) & $36_{-08}^{+08}$ & (27/75) & 24 & (18/75) & $25_{-07}^{+07}$ & (19/75) \\
\multicolumn{1}{c|}{$-$3.0} & 33 & (24/73) & $37_{-08}^{+08}$ & (27/73) & 25 & (18/73) & $26_{-07}^{+08}$ & (19/73) \\
\multicolumn{1}{c|}{$-$3.5} & 49 & (21/43) & $53_{-11}^{+11}$ & (23/43) & 40 & (17/43) & $42_{-10}^{+11}$ & (18/43) \\
\multicolumn{1}{c|}{$-$4.0} & 81 & (13/16) & $81_{-14}^{+10}$ & (13/16) & 81 & (13/16) & $81_{-16}^{+12}$ & (13/16) \\
\multicolumn{1}{c|}{$-$4.5} & 100 & (6/6) & $100_{-14}^{+00}$ & (6/6) & 100 & (6/6) & $100_{-15}^{+00}$ & (6/6) \\
\multicolumn{1}{c|}{$-$5.0} & 100 & (5/5) & $100_{-17}^{+00}$ & (5/5) & 100 & (5/5) & $100_{-17}^{+00}$ & (5/5) \\
\hline
\multicolumn{9}{c}{Stellar Streams} \\
\hline
\multicolumn{1}{c|}{$-$2.0} & 13 & (9/68) & $17_{-07}^{+08}$ & (12/68) & 1 & (1/70) & $6_{-04}^{+05}$ & (4/70) \\
\multicolumn{1}{c|}{$-$2.5} & 29 & (8/28) & $31_{-13}^{+14}$ & (9/28) & 3 & (1/30) & $13_{-10}^{+12}$ & (4/30) \\
\multicolumn{1}{c|}{$-$3.0} & 30 & (3/10) & $30_{-20}^{+24}$ & (3/10) & 9 & (1/11) & $18_{-16}^{+22}$ & (2/11) \\
\multicolumn{1}{c|}{$-$3.5} & \nodata & \nodata & \nodata & \nodata & \nodata & \nodata & \nodata & \nodata \\
% \multicolumn{1}{c|}{$-$4.0} & \nodata & \nodata & \nodata & \nodata & \nodata & \nodata & \nodata & \nodata \\
% \multicolumn{1}{c|}{$-$4.5} & \nodata & \nodata & \nodata & \nodata & \nodata & \nodata & \nodata & \nodata \\
% \multicolumn{1}{c|}{$-$5.0} & \nodata & \nodata & \nodata & \nodata & \nodata & \nodata & \nodata & \nodata
\enddata
\end{deluxetable*}

%% file: tables/tab5_dtrans.tex
\begin{deluxetable*}{llrrrrrl}[!t]
% \label{tab:dtrans}
% \tabletypesize{\scriptsize}
\tablecaption{\normalsize
	Stars within or near the ``forbidden zone'' ($D_{\rm trans} \lesssim -3.5$), which marks the threshold for efficient fine-structure line cooling in early star formation, and stars below the threshold for efficient ``standard'' dust cooling ($\rm{[Si/H]} \lesssim -4.4$) or ``shock'' dust cooling ($\rm{[Si/H]} \lesssim -5.2$).
}
\tablehead{
	\colhead{Star name} & 
	\colhead{System} & 
	\colhead{$\rm{[Fe/H]}$} & 
	\colhead{$\rm{[Si/H]}$} & 
	\colhead{$\rm{[C/H]}_{\rm corr}$} & 
	\colhead{$D_{\rm trans}^{\rm -}$} & 
	\colhead{$D_{\rm trans}^{\rm +}$} & 
	\colhead{Reference}
}
\startdata
\multicolumn{8}{c}{``Forbidden Zone'' Dust-Cooled Stars and Candidates} \\
\hline
Scl\_002\_06                 & Sculptor & $-$3.45 & $-$3.01  & $-$4.12  & $-$3.92  & $-$3.38  & \citet{Jablonka2015a} \\
Gaia~DR3~4348475068024891776 & SASS     & $-$3.56 & $-$2.91  & $-$4.10  & $-$3.90  & $-$3.36  & \citet{Hughes2026a} \\
SMSS\_J184226.25$-$272602.7  & SASS     & $-$3.89 & $-$2.95  & $-$4.09  & $-$3.89  & $-$3.35  & \citet{Jacobson2015a}; \\&&&&&&&Yelland et al. (2026b, in prep.)\tablenotemark{$\dagger$} \\
CS~30336$-$049               & SASS     & $-$4.04 & $-$3.81  & $-$3.94  & $-$3.74  & $-$3.20  & \citet{Lai2008a} \\
LMC-119                      & LMC      & $-$4.15 & $-$3.37  & $<-$3.90 & $<-$3.70 & $<-$3.16 & \citet{Chiti2024a} \\
CD$-$38$^{\circ}$~245        & SASS     & $-$4.20 & $-$3.95  & $<-$4.16 & $<-$3.96 & $<-$3.42 & \citet{Cayrel2004a} \\
J0715$-$7334                 & LMC      & $-$4.53 & $-$4.47  & $<-$3.94 & $<-$3.74 & $<-$3.20 & \citet{Ji2026a} \\
SDSS~J102915.14+172927.9     & Atari    & $-$4.97 & $-$4.26  & $<-$4.23 & $<-$4.03 & $<-$3.49 & \citet{Caffau2011d} \\
\hline
\multicolumn{8}{c}{Stars below the Dust-Cooling Efficiency Thresholds} \\
\hline
HE~1424$-$0241               & Atari  & $-$4.01 & $-$4.92  & $<-$3.17 & $<-$2.97 & $<-$2.43 & \citet{Cohen2007a} \\
J0715$-$7334                 & LMC    & $-$4.53 & $-$4.47  & $<-$3.94 & $<-$3.74 & $<-$3.20 & \citet{Ji2026a} \\
HE~0557$-$4840               & SASS   & $-$4.73 & $-$4.85  & $-$3.13  & $-$2.93  & $-$2.39  & \citet{Ji2014a}; \\&&&&&&&\citet{Norris2007a} \\
SD~1313$-$0019               & SASS   & $-$5.00 & $<-$4.79 & $-$2.02  & $-$1.82  & $-$1.28  & \citet{Frebel2015a} \\
HE~0107$-$5240               & SASS   & $-$5.54 & $<-$5.50 & $-$1.51  & $-$1.31  & $-$0.77  & \citet{Ji2014a}; \\&&&&&&&\citet{Aguado2022a} \\
HE~1327$-$2326               & SASS   & $-$5.71 & $<-$5.40 & $-$1.53  & $-$1.33  & $-$0.79  & \citet{Ji2014a}; \\&&&&&&&\citet{Frebel2008a} \\
SMSS~1605$-$1443             & SASS   & $-$6.26 & $<-$5.42 & $-$2.30  & $-$2.10  & $-$1.56  & \citet{Nordlander2019a}
\enddata
\tablenotetext{$\dagger$}{Throughout this paper, we adopt a new corrected carbon measurement for this star based on recent observations with a detection in the CH band ($A({\rm C}) = 3.70$), superseding the upper limit reported by \citealt{Jacobson2015a}; all other chemical abundances for SMSS~J184226.25$-$272602.7 are taken from \citealt{Jacobson2015a}.}
\end{deluxetable*}

%% file: appendix.tex
In the following, we provide additional details on a number of individual systems with respect to relevant observational information, notable elemental abundance features, and alternative sample sizes that further informs our CEMP fraction calculations.
\medskip

%=================================================
\section{Ultra-faint Dwarf Galaxies} \label{sec:appendix_UFD}

% ---------------
\textbf{Bootes\,I (Boo\,I)} has the largest number of CEMP stars (8 stars) and largest number of available carbon abundance measurements (36 stars) in our UFDs sample, consisting of both intermediate \citep{Norris2010c, Lai2011a} and high-resolution data \citep{Gilmore2013a, Norris2010b, Frebel2016b, Ishigaki2014a}.
Two of the four most carbon-enhanced UFD stars reside in Boo\,I; however, they border the CEMP-$s$/CEMP-$no$ region in Fig~\ref{fig:feh_cfe}, following the \citet{Yoon2016a} classification system.
Given that no neutron-capture abundances are available for a full classification and their carbon-enhancements are not extreme, these two stars cautiously remain in our sample.
The CEMP star fraction of Boo\,I is $22\%_{-10}^{+11}$ (8 CEMP stars of 36) for \feh\ $\le -2.0$, which aligns with the UFD category's average.

% ---------------
\textbf{Bootes\,II (Boo\,II)} contains four very metal-poor stars with $-3.0 \le$ \feh\ $\le -2.6$ \citep{Ji2016a}.
While their carbon abundances follow the overall Milky Way halo trend, none are carbon-enhanced.
This makes Boo\,II stand out from most other UFDs in terms of the CEMP fraction.

% ---------------
\textbf{Reticulum II (Ret\,II)} shows significant \rproc\ enhancement across most of its stars \citep{Ji2016d, Roederer2016a, Ji2018a}, making it the first known \rproc\ galaxy. 
Even so, the carbon abundances of these stars show a variety of levels, with two of the nine stars being observed as carbon-enhanced, yielding a MC CEMP star fraction of $28\%_{-25}^{+28}$ (3 CEMP stars of 8).

% ---------------
\textbf{Pictor\, II (Pic\,II)} has only one star with a carbon measurement in our sample; yet it has an upper limit of \feh\ $\le -4.6$ \citep{Chiti2025a}, classifying it at least as ultra metal-poor \citep{Chiti2025a}.
Its corresponding lower limit of \cfe\ $\ge 3.21$ is the highest value observed among the UFD systems, making it the most carbon-enhanced star in this category.
Despite this, the presence of only a single measurement prevents us from placing any meaningful constraints on CEMP fraction of Pictor\,II beyond stating that the CEMP fraction is non-zero.

% ---------------
\textbf{Pisces\, II (Pis\,II)} also has only a single star with a measured carbon abundance. 
This star is sufficiently enhanced to be classified as CEMP-$no$; however, as with Pictor\,II, it is difficult to place any meaningful constraints on the CEMP fraction of the system.

% ---------------
\textbf{Tucana III (Tuc\,III)} still has uncertainty whether it is a UFD or a globular cluster \citep{HansenT2017a, Simon2017a, Marshall2019a}. 
Based on the metallicity spread and morphological properties identified by \citet{Simon2017a}, we here consider Tuc\,III as a UFD. 
Among the limited number of stars observed, none show carbon enhancement, in contrast to the general behavior of UFD systems.
Given that CEMP stars are rare in globular clusters, this apparent absence weakens the case for a UFD classification. 
At the same time, the small number of available carbon abundance measurements unfortunately prevents any definitive distinction between these two origin scenarios. 
If more carbon abundances are obtained at lower metallicities (\feh\ $<-3.0$), then the nature of Tuc\,III could be further constrained based on the observed carbon trends of UFDs and those of globular clusters (see Section~\ref{sec:appendix_SS}).

%=================================================
\section{Classical Dwarf Spheroidal Galaxies} \label{sec:appendix_CDW}

% ---------------
\textbf{Sagittarius (Sgr)} is one of the most recent accretion events to the Milky Way, with ongoing tidal disruption and two prominent stellar streams winding around the Milky Way \citep{Majewski2003a}.
It contains only five observed CEMP stars of the 136 available carbon abundances, resulting in a MC CEMP fraction of $5\%_{-03}^{+04}$ (7 CEMP stars of 136).
\citet{Sestito2024b} and \citet{Ou2025a} have both recently measured carbon abundances in the Sgr core and streams and have found that the average is similar to that of other CDWs \citep[see also][]{HansenC2018a, Sestito2024d}.
Here, we find the median \cfe\ of Sagittarius (with 16th and 84th percentiles) is $0.04_{-0.33}^{+0.34}\,{\rm dex}$.
This is $\sim0.3$\,dex lower than the Milky Way halo at $0.31_{-0.36}^{+0.34}\,{\rm dex}$ for stars with \feh\ $\le-2.0$, though within uncertainty.

% ---------------
\textbf{Sculptor (Scl)} contains the second-largest number of carbon abundances (80 measured and 12 upper limits) and the largest number of observed CEMP stars (14 stars) in our CDW sample, resulting in a MC CEMP fraction of $16\%_{-06}^{+07}$ \citep{Chiti2018a, Jablonka2015a, Simon2015a, Skuladottir2015a, Skuladottir2024a, Starkenburg2013a, Kirby2012c, Frebel2010b}.
The median \cfe\ abundance of Sculptor is $0.35_{-0.42}^{+0.43}\,{\rm dex}$, consistent with that of the Milky Way halo and higher than that of Sagittarius by $\sim0.3\,{\rm dex}$.

%=================================================
\section{Accreted Dwarf Galaxies} \label{sec:appendix_ADW}

% ---------------
\textbf{Atari Disk (Atr)}, a recently recognized population of metal-poor stars, is reminiscent of the ``metal-weak thick disk'' \citep{Norris1985a, Mardini2022a, Mardini2024a}.
For this work, the original Atr sample of \citet{Mardini2022a} has been redefined following the same procedure applied to our other systems (see Section~\ref{sec:lit_prep}). 
Stars exhibiting extrinsic carbon enrichment (from the \sproc\ and \iproc) and those having abundances outside of the adopted abundance ranges were removed accordingly.
We additionally excluded all confirmed Atr stars from our Milky Way halo sample \citep{Placco2014c}, as many of these objects were previously classified as halo stars.

Overall, Atr members display a moderate level of carbon enhancement across the entire low-metallicity range.
The CEMP fraction is broadly consistent across uncertainties with levels of $57\%_{-30}^{+28}$ for \feh\ $\le -3.5$, $33\%_{-31}^{+40}$ for \feh\ $\le -4.0$, and $0\%_{-00}^{+50}$ for \feh\ $\le -4.5$ (of one measurement).
Notably, Atr contains a significant number of the most iron-poor stars (\feh\ $\le -4.0$), albeit with a low CEMP fraction, consistent with Atr being one of the oldest dynamically coherent components of the proto–galactic disk \citet{Mardini2022a}.

When comparing these carbon abundances and CEMP fractions to the GSE, we find strong differences though they are both fully accreted systems (see Figure~\ref{fig:feh_cfe}).
This suggest that Atr likely follows a different pre-accretion history than the GSE and the classical dwarfs, as discussed Section~\ref{sec:supernovae}.

% ---------------
\textbf{Gaia-Sausage/Enceladus (GSE)} is another collection of stars believed to be the merger debris of a disrupted dwarf galaxy, and thought to have accreted onto the Milky Way $t_{\rm acc}=9.1^{+0.7}_{-0.7},\mathrm{Gyr}$ ago \citep{Helmi2018a, Belokurov2018b, Kruijssen2020c}.
The GSE sample was taken from \citet{Ou2024c}, but only limited abundance data exist in the low metallicity regime, and no extremely metal-poor stars have been identified to date.
As a result, firm conclusions on the CEMP fraction remain difficult to draw.
Given that GSE is predicted to share a similar pre-accretion evolutionary history with classical dwarfs, the absence of carbon-enhanced stars broadly aligns with we find for the classical dwarf galaxies.

% ---------------
\textbf{Large Magellanic Cloud (LMC)} is a massive, gas-rich dwarf irregular galaxy currently in the process of being accreted by the Milky Way. 
Despite having one of the largest samples of stars with measured carbon abundances (26 stars), none of the stars have been observed with \cfe\ $\ge 0.7$ \citep{Chiti2024a, Ji2026a}.
That said, through our statistical analysis, we present a non-zero MC CEMP fraction for the LMC due to the proximity of four stars very close to the carbon-enhancement threshold.
This overall lack of carbon enhancement is similar to that observed in the GSE and reminiscent of the classical dwarf galaxies, suggesting that the LMC may have experienced a similar formation and pre-accretion evolutionary history. 
We also note that five stars have recently been identified as CEMP stars in the LMC \citep{Lucey2026a}, but they have been excluded from our sample by means of the \citet{Yoon2016a} classification given that their carbon enhancements are likely due to extrinsic enrichment (CEMP-$s$). 
However, their neutron-capture abundances are yet to be determined for final confirmation.

% ---------------
\textbf{``Small Accreted Stellar System'' (SASS)} stars represent a chemically and kinematically distinct population of extremely metal-poor stars \citep{Andales2024a, Hughes2026a}.
The defining characteristic of SASS stars are their extremely low neutron-capture abundances, specifically in strontium ($\rm{Sr/H} \le -4.5$) and barium ($\rm{Ba/H} \le -4.0$), which are nearly four orders of magnitude below the typical halo abundances.
Notably, the SASS sample contains several of the most chemically primitive stars known, such as HE~0107$-$5240 ($\rm{[Fe/H]} = -5.5$; \citealt{Aguado2022a}), HE~1327$-$2327 ($\rm{[Fe/H]} = -5.7$; \citealt{Frebel2008a}), SMSS~J160540.18$-$144323.1 ($\rm{[Fe/H]} = -6.2$; \citealt{Nordlander2019a}), and 2MASS~J03130036$-$6708393 ($\rm{[Fe/H]} < -7.3$; \citealt{Keller2014a}).
Of the 78 stars currently classified as SASS stars, 77 have measured carbon abundances (including two with high upper limits).
Among these, 24 have been observed as carbon-enhanced, corresponding to a MC CEMP fraction of $36\%_{-08}^{+08}$ (27 CEMP stars of 76); the highest among all systems analyzed in this study.
Many of these stars were previously included in the Milky Way halo sample from \citet{Placco2014c}, but we have reclassified here to avoid duplication.

%=================================================
\section{Stellar Streams} \label{sec:appendix_SS}

%---------------
\textbf{C-19} is the most metal-poor (disrupted) cluster known thus far \citep{Martin2022a}.
Six of the eight stars have measurable carbon abundances, and none of them are considered carbon-enhanced.
Interestingly, these six stars occupy a very small metallicity range of $-3.15 \le$ \feh\ $\le -3.41$, well below the current metallicity floor for globular clusters.
Together, this may suggest that the C-19 stream progenitor formed from gas enriched differently from that of typical globular clusters or even other extremely metal-poor halo stars with higher \cfe\ values.

%---------------
\textbf{Helmi stream} was the first stellar discovered stream through kinematic phase space analysis \citep{Helmi1999b}.
It has been has been further decomposed into two parts \citep{Beers2017a}: the Helmi debris stream and the Helmi trail stream, but here, we jointly consider the debris and trail streams together.
Relative to the other stellar streams, the Helmi stream spans the largest metallicity range from $-3.4 <$ \feh\ $< -2.1$ and has the most diverse carbon abundances \citep{Gull2021a, Roederer2010a}.
Though the progenitor of the Helmi stream is still not known, the trend in carbon abundances and carbon enhancement is more consistent with that of the ultra-faint dwarf galaxies, rather than a globular cluster.

%---------------
\textbf{Omega Centauri stream} is associated with tidal debris from the massive globular cluster $\omega$~Centauri, which is often interpreted as the remnant nucleus of a disrupted dwarf galaxy \citet{Bekki2003a, Majewski2012a}.
For the ten stars in our sample, the $\omega$-Centauri stream spans a relatively narrow metallicity range of $-2.6 \le$ \feh\ $\le -2.0$, yet exhibits a comparatively wide diversity of carbon abundances, spanning $-0.3 \lesssim$ \cfe\ $\lesssim 1.0$ \citep{Gull2021a}.
This \feh\ range is consistent with its association to a globular cluster progenitor, whereas the broader spread in \cfe\ more closely resembles that of dwarf galaxy environments.
Although the current sample size is limited, the observed carbon abundance distribution provides tentative chemical evidence in favor of a dwarf galaxy origin.
Additional carbon measurements will be critical for more robustly constraining the nature of the progenitor.
Based on the presently observed abundances, we adopt a dwarf galaxy origin for $\omega$~Centauri for comparisons with other stellar stream systems.